\documentclass[english, 12pt]{article}

\usepackage{amsfonts}
\usepackage[T1]{fontenc}
\usepackage{amsmath}
\usepackage{graphicx}
\usepackage{babel}
\usepackage{xcolor}
\usepackage{amssymb}
\usepackage[authoryear]{natbib}

\newcommand{\bb}[1]{\boldsymbol{#1}}
\newcommand{\Letteranuova}{\Omega}
\newcommand{\letteranuova}{a}

\newcommand{\taunew}{\tau ^\star}
\newcommand{\mc}[1]{\mathcal{#1}}

\title{Learning in Associative Networks through Pavlovian Dynamics}
\date{}
\author{}
\begin{document}
\maketitle

{\bf \large  Daniele Lotito$^{1,2}$,  Miriam Aquaro$^{2,3}$,  Chiara Marullo$^{2,4}$}
\\ 
{$^{\displaystyle 1}$Dipartimento di Informatica, Universit\`a di Pisa, Pisa, Italy}\\
{$^{\displaystyle 2}$GNFM-INdAM, Gruppo Nazionale di Fisica Matematica, Istituto Nazionale di Alta Matematica, Italy}\\
{$^{\displaystyle 3}$Dipartimento di Matematica, Sapienza Universit\`a di Roma, Roma, Italy}\\
{$^{\displaystyle 4}$National Research Council (CNR), Institute for High Performance Computing and Networking (ICAR), Napoli, Italy}


\thispagestyle{empty}
\markboth{}{NC instructions}
\ \vspace{-0mm}\\
\begin{center} {\bf Abstract} \end{center}

Hebbian learning theory is rooted in Pavlov's Classical Conditioning.\\ While mathematical models of the former have been proposed and studied in the past decades, especially in spin glass theory, only recently it has been numerically shown that it is possible to write neural and synaptic dynamics that mirror Pavlov conditioning mechanisms and also give rise to synaptic weights that correspond to the Hebbian learning rule.

In this paper, we show that the same dynamics can be derived with equilibrium statistical mechanics tools and basic and motivated modeling assumptions.
Then, we show how to study the resulting system of coupled stochastic differential equations assuming the reasonable separation of neural and synaptic timescale. In particular, we analytically demonstrate that this synaptic evolution converges to the Hebbian learning rule in various settings and compute the variance of the stochastic process.

Finally, drawing from evidence on pure memory reinforcement during sleep stages, we show how the proposed model can simulate neural networks that undergo sleep-associated memory consolidation processes, thereby proving the compatibility of Pavlovian learning with dreaming mechanisms.


%

\section{Introduction}
Hebbian learning theory has been extensively studied in various fields, and the mathematical implementation of Hebb's rule is crucial in many spin glass models, the most famous being Hopfield model~\citep{AGS-PRL}.
Hebb's rule is usually stated as ``neurons that fire together wire together''~\citep{Shatz1992} or, more formally, ``any two cells or systems of cells that are repeatedly active at the same time will tend to become `associated' so that activity in one facilitates activity in the other''~\citep{Cooper2013}.
However, what Hebb wrote in The Organization of Behaviour~\citep{Hebb} is ``when an axon of cell A is near enough to excite a cell B and repeatedly or persistently takes part in firing it, some growth process or metabolic change takes place in one or both cells such that A's efficiency, as one of the cells firing B, is increased''.
The most known statement differs from the original one in the sense that the former is based on mere contiguity, while the latter includes temporal precedence, causality and contingency~\citep{Keysers2014}.
The mathematical expression of Hebb's rule in the Hopfield model is so straightforward that it could be considered derived from either statement. This is evident in the Hebbian prescription for the coupling matrix $J_{ij} \propto \sum_{\mu = 1}^{K} \xi _i ^\mu \xi _j^\mu$, where the matrix element $J_{ij}$  models the association between neuron $i$ and neuron $j$, and this element is proportional to the number of patterns $\{\boldsymbol{\xi}^\mu\} _{\mu = 1,2,..., K}$ fed to the network whose $i$ and  $j$ entries assume the same value.
 
However, when modeling neurophysiological phenomena such as Pavlov's classical conditioning, the simplification ``neurons that fire together wire together'' falls short~\citep{Keysers2014}, since in classical conditioning the stimuli can be used to predict the occurrence of other stimuli ~\citep{clark2004classical}.
Moreover, Hebb frequently references Pavlov's theory, and the learning process he describes closely aligns with Pavlov's behavioral approach, reflecting the influence of early behaviorist ideas in Hebb's work.\footnote{ The MA thesis of Hebb is entitled ``Conditioned and Unconditioned Reflexes and Inhibition'', while his PhD supervisor was the american psychologist and behaviorist, Karl Lashley~\citep{milner2003brief, langille2018synaptic}.}
Nonetheless, a rigorous mathematical model of Pavlov's conditioning that accounts for temporal precedence, causality, and contingency can be constructed. Recent work in~\citep{agliari2022pavlov} illustrates that systems of Ising neurons, influenced by an external field, provide a meaningful mathematical representation of traditional Pavlovian conditioning, and numerical experiments show that the related learning process leads to the Hebbian prescription for storing network memories. In particular, the network dynamically learns memories and stores them in a coupling matrix whose function is coordinating the neural activity and which is equal to the coupling matrix of the Hopfield model, deeply studied in statistical mechanics. This equality is not analytically characterized in~\citep{agliari2022pavlov}, but its validity is corroborated with numerical experiments.  \\
It is thus natural to ask whether (1) we can obtain the same Pavlovian dynamics introduced in~\citep{agliari2022pavlov} within the statistical mechanics framework, (2) what analytical conclusions can be drawn about the convergence to the Hebbian coupling matrix, and (3) how versatile this model is.\\

Indeed, in this case, with the basic modeling assumptions, we write the partition function for a network of neurons and synapses that allows us to set up the ODE system that describes the coupled evolution of neurons and synapses. This leads to a mathematical model of Pavlov's Classical Conditioning using spin glass theory tools, which we will outline briefly in Sec.~\ref{sec:SM}.\\
In particular, we find that the separation of neural and synaptic time-scale corresponds to conditioning on the synapses when computing the Boltzmann averages
as if the synaptic and the neural subsystems need to be ``adiabatically" separated. Since we are using tools of equilibrium statistical mechanics we need to postulate the relaxation dynamics towards the equilibria, and we chose an exponential decay. 
We then analytically prove its convergence to the Hebbian kernel, a result that was only numerically supported in the aforementioned work. Subsequently, we analyze settings that more closely mirror what happens in nature, in which the stimuli presented to the neurons are not chosen from the uniform distribution across possible stimuli.\\

Once having proved that this Pavlovian model for associative memory implies Hebbian learning, we argue that this model is suited for describing biological networks.
When modeling biological learning networks, we need to check the robustness of the resulting model with respect to the fundamental features of biological learning.  One key feature is sleep, which, as biological evidence suggests, promotes the reactivation of neural circuits involved in learning, thereby strengthening synaptic connections related to newly acquired information~\citep{crick1983function}. This process, often referred to as memory replay or neuronal replay, occurs predominantly during specific stages of sleep, such as slow wave sleep (SWS) and rapid eye movement sleep (REM). In particular, these two stages of sleep play a critical function in filtering and prioritizing memories, which helps prevent cognitive overload and ensures efficient storage of relevant information. This selective retention mechanism involves strengthening important memories while weakening or discarding irrelevant or spurious ones, a process essential for optimal cognitive functioning. This led to the development of computational models that aimed to mimic these biological learning dynamics~\citep{Hassabis2017}.

In the associative neural networks literature this function is accomplished by the dreaming mechanism, a well-established and biologically plausible way to mitigate crosstalk effects between memories~\citep{Fachechi2019, Agliari2019c, aquaro2022recurrent}.
Thus, in the spirit of describing biological networks within our framework, we have to demand that the model can describe neural networks that sleep, here we show that this is possible. The resulting network can handle larger information content and more effectively than in the Hebbian case.

The paper is structured as follows: Sec.~\ref{sec:SM} introduces the spin glass modeling of Pavlov's Classical Conditioning, which enables the formulation of a multiscale system of ordinary differential equations representing neural and synaptic Pavlovian dynamics. 
Sec.~\ref{sec:Pavlov_implies_Hebb} discusses the convergence of the dynamics to the Hebbian prescription for the coupling matrix of associative neural networks and provides a discrete version of this Pavlovian dynamics. Sec.~\ref{sec:dreaming} generalizes the dynamics of the model, showing that, using the same mathematical tools, it is possible to derive the so-called dreaming prescription for the coupling matrix.
Sec.~\ref{sec:experiment} is dedicated to numerical checks and experiments analyzing the convergence of the coupling matrix under different scenarios where the stimuli are not uniformly selected. 
Finally, Sec.~\ref{sec:end} is left for conclusions and outlooks.


\section{A statistical mechanics view on neural and synaptic Pavlovian dynamics} \label{sec:SM}
In this section, we introduce the statistical mechanics formulation of Pavlov's classical conditioning, a more formal derivation is present in~\citep{agliari2022pavlov}, and it is discussed in this paper in Appendix~\ref{sec:appA}. \\
We consider a system composed of \(N\) binary neurons whose configurations are described by $\boldsymbol{\sigma}\in\Sigma_N:=\{-1,1\}^N$.
The system's thermodynamics is characterized by the following 2-point Hamiltonian:
\begin{equation} \label{eq:H}
H_{N}(\bb \sigma,\bb J) = -\frac{1}{2}\sum_{i\neq j=1}^{N,N}J_{ij}\sigma_{i}\sigma_{j}-u\sum_{i=1}^N h_{i}\sigma_{i},
\end{equation}
where \(\bb J \in\mathbb{R}^{N\times N}\) represents the symmetric coupling matrix describing pairwise neuron interactions, \(h_i \in \{-1,0,1\}\) is the bias or, in more biological terms, the {\em firing threshold} acting on each neuron and \(u\) is a global amplification factor that modulates the strength of this bias.
Let us now extend the configuration space to \(\Sigma_N\times \mathcal{J}_{N}\), where $\mathcal{J}_{N}$ represent the space of the $N\times N$ symmetric matrices with null diagonal elements and equip this space with a Boltzmann-Gibbs measure
\begin{equation}\label{PNB}
\mc P_{N}(\beta, \bb \sigma,\bb J)=\frac1{Z_{N} (\beta)} \exp\big[-\beta H_{N}(\bb\sigma,\bb J)\big],
\end{equation}
where
\begin{equation}\label{PartitionFunction}
Z_{N}(\beta)=\sum_{\bb J\in \mathcal{J}_{N}} \sum_{\bb\sigma\in \Sigma_N} \exp\big[-\beta H_{N}(\bb \sigma,\bb J)\big]
\end{equation}
is the associated partition function. The parameter \(\beta\) governs the broadness of the distribution, dictating the noise level affecting the neuron dynamics. As \(\beta \to 0\), the distribution becomes flat, and each neuron is randomly oriented. Conversely, as \(\beta \to \infty\), the distribution sharply peaks at the configurations that minimize the Hamiltonian.
In the classical statistical mechanics framework applied to neural networks, such as in pattern recognition applications~\citep{amit1989,CKS}, synapses are considered as \emph{quenched} variables, meaning that they remain fixed. Consequently, the summation over \(\bb J \in \mathcal{J}_{N}\) will not appear in the definition of the partition function~\eqref{PartitionFunction}. However, in this work, we aim to explore the coupled dynamical evolution of neurons \(\boldsymbol{\sigma}\) and synapses \(\boldsymbol{J}\). This scenario requires that we do not treat the couplings as quenched variables. Instead, the partition function should now include the summation over synapse configurations,  and this corresponds to the so-called \emph{annealed expression}~\citep{MPV}.

\subsubsection*{Pavlov, Hebb and statistical mechanics}
Let us now discuss how this picture can be related to Pavlov's conditioning.
\footnote{We recommend \citep{clark2004classical} for a historical perspective on Pavlov conditioning.}
At its core, Pavlovian (or classical) conditioning involves learning to associate a neutral stimulus with a naturally occurring stimulus, leading the neutral stimulus to evoke a learned response.
We first observe that typical thermalization processes of Hopfield-like models~\citep{hopfield1982hopfield} already partially suit this modeling purpose. In these models, if the initial configuration is similar to a stored pattern, the equilibrium state will be the pattern itself.
The stored pattern is usually a binary vector, and let us view this vector as the concatenation of two vectors.
One of them corresponds to the so-called \textit{Unconditioned Stimulus}, which is the stimulus that in Pavlovian conditioning naturally and automatically triggers the \textit{Unconditioned Response}; in the famous salivation experiment, food is the US, which causes the dog to salivate.
The other one corresponds to the \textit{Conditioned Stimulus},  a previously neutral stimulus that does not initially trigger a specific response. For example, a bell sound in Pavlov’s experiments.
If the initial configuration is the Conditioned Stimulus, and the stored pattern is the concatenation of both, the relaxation process towards thermodynamic equilibrium will consist of the retrieval of the Unconditioned Stimulus. In turn, it will trigger the response, and, since in this case the Unconditioned Stimulus is not directly supplied to the network,  this response is called \textit{Conditioned Response}.
The missing piece in this picture is the pairing between the two stimuli, which should be created by repeated paired stimulation -- Pavlov's prescription-- and 
not directly supplied to the network by setting the coupling matrix --Hebbian prescription.\footnote{A note on the terminology, with Hebbian prescription we mean the Hopfield model form of the coupling matrix. As we discussed in the introduction, reducing Hebb's theory to the mathematical form of the coupling matrix is an oversimplification. In this work, we are trying to highlight the connection between Pavlovian and Hebbian pictures from a mathematical modeling perspective.}
From a modeling perspective, we feel that a proper explanation of the mathematical link between these two phenomena is still missing. While~\citep{agliari2022pavlov} does an excellent job at presenting numerical experiments for the joint retrieval of the paired stimuli starting from a single one (an interesting case discussed therein is the coupling between more than two different stimuli), here we thoroughly discuss the derivation neural and synaptic dynamics rules from an equilibrium statistical mechanics treatment and modeling considerations.\\
We discuss why this dynamics is linked to both Pavlov classical conditioning and Hebbian learning and study the analytical convergence of this dynamics to the Hebbian prescription
both in the (unrealistic) case in which patterns are presented cyclically and in the (more realistic) case in which patterns are presented according to a stochastic process. In this case, we also compute the variance of the process. \\

We started from the Hamiltonian~\eqref{eq:H}, and obtained the Boltzmann-Gibbs measure~\eqref{PNB}, we have written explicitly the dependency on the couplings $\bb J$ so that we can compute its average with respect to this measure. This quantity  depends on the spin averages with respect to this measure. Moving to the spin activity average, we first assume that the neural subsystem evolves faster than the synaptic one so that we can compute the average conditioning on the coupling realization, i.e. keeping $\bb J$ fixed.\\
Once we find the equilibria, we write a relaxation process towards them, guided by Hebb's theory and common practices.






\par\medskip
\medskip\par
\subsubsection*{Building a Pavlovian dynamics}
We now consider an interaction matrix space characterized by the Rademacher probability measure, i.e.
\begin{equation}\label{eq:rademacher}
P(J_{ij}=\pm1)=1/2, \quad \text{for} \ i,j = 1, \dots, N, \quad i \neq j.
\end{equation}
In other words, we are choosing a coupling matrix whose entries are i.i.d. random variables drawn from a Rademacher distribution.
To simplify computations, we add in the partition function from~\eqref{PartitionFunction} 
the two source terms \(\bb t\) and \(\bb T\) for neurons and weights respectively
\begin{equation}\label{starting}
{Z}_{N}(\beta, \bb t, \bb T)=\sum_{\bb \sigma}\sum_{\bb J}\exp\Big(\beta\sum_{i<j}^N J_{ij}\sigma_{i}\sigma_{j}+\beta u \sum_{i=1}^N h_{i}\sigma_{i}+\sum_{i=1}^N t_{i}\sigma_{i}+\sum_{i<j}^NT_{ij}J_{ij}\Big),
\end{equation}
where the sum over the entries of the coupling matrix is explicitly given by
\[
\sum_{\bb J} \equiv \prod _{i<j} \sum_{J_{ij}=\pm1}.
\]
With a slight abuse of notation, we denote ${Z}_{N}(\beta, \bb t, \bb T ; \bb J)$ the partition function relative to the smaller configuration space $\Sigma _N$. This accounts for the case in which the couplings $\bb J$ are treated as quenched variables, meaning they are considered fixed parameters. The analytical expression of ${Z}_{N}(\beta, \bb t, \bb T ; \bb J)$ differs from eq.~\eqref{starting} because there is no summation over the couplings, see eq.~\eqref{eq:starting_J}. \\
Summing over all \(J_{ij}\) variables in eq.~\eqref{starting} results in
\begin{equation}
{Z}_{N}(\beta, \bb t, \bb T)=\sum_{\bb\sigma}\exp\Big(\sum_{i=1}^N\left(\beta u h_{i}+t_{i}\right)\sigma_{i}\Big)\prod_{i<j}2\cosh\left(\beta\sigma_{i}\sigma_{j}+T_{ij}\right). \label{eq:inducingZ}
\end{equation}
By deriving with respect to the source \(T_{ij}\) and then setting both the source terms to zero, we find, see Appendix~\ref{sec:app0} for details,
\begin{equation} \label{eq:a}
\langle J_{ij}\rangle=\frac{\partial\log {{Z}_{N}(\beta, \bb t, \bb T)}}{\partial T_{ij}}\Big\vert_{\bb T=0, \bb t =0}=\langle\sigma_{i}\sigma_{j}\rangle\tanh\beta \approx \langle\sigma_{i} \rangle \langle \sigma_{j} \rangle  \tanh\beta,
\end{equation}
where the average is related to the measure induced by~\eqref{eq:inducingZ}. Equation \eqref{eq:a} tells us that no matter the initial value of the coupling matrix, its equilibrium value depends only on the neural average, and this descends from the fact that in the partition function~\eqref{eq:inducingZ} the coupling realization is summed out.  The last step in  Equation \eqref{eq:a} corresponds to the mean-field assumption under which each neuron is subject to a net effective external field, rather than to the interactions with other spins.
\newline
Similarly, see again Appendix~\ref{sec:app0} for details, the spin activity expectation  yields:

\begin{equation} \label{eq:b}
\langle\sigma_{i}\rangle  _{\bb J _\text{fixed}}=\frac{\partial\log {{Z}_{N}(\beta, \bb t, \bb T; \bb J)}}{\partial t_i}\Big\vert_{\bb T=0, \bb t =0}\approx\tanh\Big(\beta\sum_{j\neq i}^N J_{ij}\langle\sigma_{j}\rangle _{ \bb J _\text{fixed}}+\beta u h_{i}\Big) . 
\end{equation}
The requirement of keeping the coupling fixed amounts to assume that relaxation processes of neurons and synapses happen in different time scales, a fact that is true in biological systems~\citep{maass1997dynamic}. The fact that we need to condition on a specific coupling realization corresponds to assuming that the neural equilibrium is reached faster than the synaptic one: we let the neural subsystem relax to its stationary state every time the couplings vary. The relationship between the hierarchical time scale structure and statistical mechanics averaging schemes is also discussed in~\citep{Coolen_2001, Mourik_2001}.
We compute Eq.~\eqref{eq:a} and Eq.~\eqref{eq:b} in appendix~\ref{sec:app0}.\footnote{We also comment that an alternative approach could be sampling configurations of neurons and couplings directly from the Boltzmann measure (conditioned on the coupling realization for the case neural variables) with a Monte Carlo strategy. Montecarlo sampling can even be used after the mean-field approximation to sample the average spin configuration directly from the Boltzmann measure. In this work, we choose to pursue the analytical route, especially because sampling from the joint distribution of neurons and synapses requires overcoming computational challenges. 
Moreover, since we aim for a model in which the dynamics evolution is determined for each neuron individually, the mean-field approximation is a natural choice. This approximation simplifies the analysis by averaging the effects of interactions among neurons. As we will see, this approach seamlessly leads to the Hebbian kernel, which characterizes models where every neuron interacts with all others for which the mean-field approximation is particularly appropriate.
}

Given the equivalence of all spin indices, the results from Eqs. (\ref{eq:a}) and (\ref{eq:b}) remain consistent for all \(i\) and \(j\), with \(i \neq j\).\newline
\medskip

We now choose the exponential decay dynamics towards the equilibria that we have derived.
This choice is motivated by its common use in studies of spin system dynamics~\citep{suzuki1968dynamics}, as well as in the modeling of neural systems~\citep{krotov2021large} and behavioral phenomena~\citep{boschi2020opinion}.
Additionally, the experimental validity of the ``converse'' of Hebb's postulate further supports this choice~\citep{ wiesel1963single, stent1973physiological}. In particular, it is supported by evidence that when the pre-synaptic axon of cell A repeatedly and persistently fails to excite the post-synaptic cell B while cell B is firing under the influence of other pre-synaptic axons, metabolic changes take place in one or both cells such that A’s efficiency as one of the cells firing B is decreased~\citep{milner2003brief}. 

 We introduce one time scale for the synapses and one for the neurons and we acknowledge that while~\eqref{eq:a} must be true at equilibrium, the evolution of the couplings can depend only on the equilibrium of the faster subsystem, thus we make the substitution  $\langle\sigma_{i} \rangle\to \langle\sigma_{i} \rangle_{\bb J _\text{fixed}}$  in \eqref{eq:b}. In order to lighten the notation we set \(\langle J_{ij}\rangle=J_{ij}\) and  \(\langle\sigma_{i} \rangle_{\bb J _\text{fixed}} = \sigma_{i}\). We thus write
\begin{align}
	\dot{{\sigma_{i}}} & =-\frac{1}{\tau}\sigma_{i}+\frac{1}{\tau}\tanh\Big(\beta\sum_{j\neq i}^{N}\ J_{ij}\sigma_{j} + \beta u h_{i}\Big),\label{eq:sigma_rad}\\
	\dot{{J_{ij}}} & =-\frac{1}{\tau^{\prime}} J_{ij}+\frac{1}{\tau^{\prime}}\sigma_{i}\sigma_{j}\tanh\beta,\label{eq:J_rad}
\end{align}
with \(\tau\) and \(\tau'\) being the neural and synaptic time-scales.  We will keep these time scales separated \(\tau \gg \tau'\),  this means that if we focus on the neural evolution we will consider the synapses fixed, according to~\eqref{eq:b}. We call this dynamics Pavlovian dynamics because can be used to express the pairing process of stimuli, as discussed before and numerically corroborated in~\citep{agliari2022pavlov}.
\footnote{
We remark that the derivation of Eq.~\eqref{eq:sigma_rad} and Eq.~\eqref{eq:J_rad} presented in this previous section is not unique; we have sketched in Appendix~\ref{sec:appA} the master equation approach used in~\citep{agliari2022pavlov}.}

\section{Pavlovian dynamics implies Hebbian learning}\label{sec:Pavlov_implies_Hebb}
In this section, we will focus on the evolution of the synaptic coupling matrix derived in the previous section, showing that the network learns the associations between patterns and converges to the Hebbian prescription. We will prove this convergence both in the case of the continuous dynamics described by the Eq.\eqref{eq:J_rad} (see Sec.~\ref{sec:continuous}) and for its discrete approximation (see Sec.~\ref{sec:discrete} and Appendix~\ref{sec:appB}).
Finally, in Sec.~\ref{sec:prob} we will present the setting in which the patterns are extracted from a given probability distribution. In this case, the coupling matrix probability distribution converges to the distribution of the asymptotic states, whose mean and variance are analytically computed.

\subsection{Convergence in the continuous case}\label{sec:continuous}
Let us start by analysing the dynamics described by Eq.~\eqref{eq:J_rad}.  Assuming that the synaptic dynamics leads to a steady state, without loss of generality, we can set \(\dot{{J_{ij}}} = 0\) thus obtaining \( J_{ij} = \tanh(\beta) \sigma_i  \sigma_j \). We can also enforce two singular-bit stimuli, \(\xi_i\) and \(\xi_j\), onto the neurons \(\sigma_i\) and \(\sigma_j\), which means that we can set \(\sigma_i = \xi_i\) and \(\sigma_j = \xi_j\). We can also directly set \(\bb \sigma = \bb\xi\); we call this procedure \textit{presenting a pattern to the network}.
We observe that in this case the synaptic matrix elements
\begin{equation}\label{eq:JJ}
J_{ij} = \tanh (\beta) \, \xi_i \xi_j \, 
\end{equation}
resembles the Hebbian prescription.
In Eq.~\eqref{eq:JJ}, the parameter \(\beta\) controls the coupling strength. When \(\beta \to 0\), the noise dominates the stimuli, preventing any learning.
Otherwise, if \(\beta \to + \infty\) we have that  $\tanh (\beta)\approx 1$ and ~\eqref{eq:JJ} corresponds to the Hebbian prescription
\begin{equation}\label{eq:Hebb_matrix_element}
J_{ij}=\frac{1}{K} \sum_{\mu =1 }^{K}\xi^{\mu}_i \xi^{\mu}_j,
\end{equation}
in case of a single pattern $K=1$.\\
We also notice that, in this simple case, we have full knowledge of the synaptic matrix dynamics. In fact, if  we fix $\bb \sigma = \bb \xi ^1 $, and denote $J(t):= J_{ij}(t), \ J_0 = J(t = 0) \ c:= \xi _i ^1 \xi _j ^1  \tanh\beta$, Eq.~\eqref{eq:J_rad} reduces to the first-order linear ordinary differential equation
\begin{equation} \label{eq:eqdiff_k1}  
\frac{dJ}{dt} = -\frac{1}{\tau'} J(t) + \frac{c}{\tau'}.
\end{equation}
The solution to the differential equation~\eqref{eq:eqdiff_k1} can then be written as 
\begin{equation}\label{eq:eqdiff_k1_sol}
J(t) = (J_0 - c) e^{-\frac{1}{\tau'}t} + c.\end{equation}
%
Thus, \(J(t)\) approaches exponentially the steady-state value \(c\), starting from \(J_0\), with a rate determined by \(\tau'\).\\
In general, $c$ is not a constant. For instance, in the case of multiple patterns  $c = c(t) = \sigma_{i} (t) \sigma_{j}(t) \tanh\beta $ varies every time a pattern is presented to the network, we explicit in the following subsection in which conditions we can effectively ``present patterns to the network" acting only on the fields $\{ h_i\}_{i = 1,\dots, N}$ and evolving the neurons with the dynamics~\eqref{eq:sigma_rad}.
In order to have $c(t)$ behave as a constant we have to replace $c$ with its temporal average over an intermediate scale $\taunew $ between the synaptic scale $\tau '$ and the neural scale $\tau$.\footnote{Formally, we are computing a temporal average of~\eqref{eq:eqdiff_k1} over this new scale and considering $J(t)$ constant during this time integral.} We thus have 
\begin{equation}\label{eq:effective_c_temporal_average}
c  \equiv \frac{1}{\taunew} \int _{t- \taunew} ^{t } \text{d} \tau \, \sigma_{i} (\tau) \sigma_{j}(\tau) \tanh\beta .
\end{equation}
Assuming that, regardless of $t$ and $\tau'$, in the temporal window $\taunew$ neurons are aligned to each pattern $\bb \xi ^ \mu$ an amount of time that is proportional to the probabilities $\{ p_\mu \}_{\mu=1,\ldots,K}$, and the separation of the time-scale $\tau \ll \taunew \ll \tau '$,  we can simply set in~\eqref{eq:eqdiff_k1_sol} $c = \sum _ {\mu =1} ^ {K}  p_\mu \xi^{\mu}_i \xi^{\mu}_j \tanh{\beta}$. Then, if $\beta \gg 1$ and  $p_\mu = \frac{1}{K}$ for each pattern, we recover the Hebbian coupling prescription ~\eqref{eq:Hebb_matrix_element}.
\subsubsection*{Hierarchy of time scales}
We stress that the separation $\tau \ll \taunew \ll \tau '$ allows that \eqref{eq:effective_c_temporal_average} is constant in the synaptic time-scale.
The procedure of presenting a pattern to the network is equivalent to assuming that the neural time scale $\tau $ is by far the shortest time scale of the problem and the neural fields $\{h_i\}_{i=1,\dots, N}$ are strong $\sum_{j\neq i}^{N}\ J_{ij}\sigma_{j} \ll u h_{i}$ and $\beta u \gg 1$, see equation~\eqref{eq:sigma_rad}. The condition on $\taunew$, $\tau \ll \taunew \ll \tau '$ with the further requirement that the strong neural fields change quickly with respect to $\taunew$ justifies the ergodic assumption \eqref{eq:effective_c_temporal_average}.


\subsection{Convergence in the discrete case}\label{sec:discrete}
Building upon the convergence results established in the preceding section, the objective now is to move towards a discrete formulation of Eqs.~\eqref{eq:sigma_rad} and~\eqref{eq:J_rad}, showing that, under different settings, the convergence to the Hebbian prescription ~\eqref{eq:Hebb_matrix_element} continues to yield.

Let us start by writing the discrete version of Eqs.~\eqref{eq:sigma_rad} and~\eqref{eq:J_rad} as in~\citep{agliari2022pavlov}:

\begin{align}
	\sigma_{i}^{(n+1)} &= \sigma_{i}^{(n)}\left(1-\frac{\delta t}{\tau}\right) + \frac{\delta t}{\tau}\tanh\left(\beta\sum_{j\neq i}^{N}J_{ij}^{(n)}\sigma_{j}^{(n)}+ \beta u  h_{i}^{(n)}\right),\\
	J_{ij}^{(n+1)} &= J_{ij}^{(n)}\left(1-\frac{\delta t}{\tau^{\prime}}\right) + \frac{\delta t}{\tau^{\prime}}\sigma_{i}^{(n)}\sigma_{j}^{(n)}\tanh\beta,
\end{align}
where \(n\) denotes the discrete time step, \(\delta t\) represents the differential time unit responsible for system evolution and \( h_i^{(n)} \)  corresponds to an external stimulus that varies over time. \\
\
Since we are mainly interested in studying the convergence of the coupling, and we discussed that \( \tau' \) is substantially larger than \( \tau \),  we choose to set \( \delta t = \tau \), that corresponds to allow the spins to align instantly with the stimulus, at least in the strong field regime $\beta u \gg 1$ which we consider.  In this way, we obtain 
\begin{eqnarray}
	\sigma_{i}^{(n+1)}&=&\tanh\big(\beta\sum_{j\neq i}^{N}J_{ij}^{(n)}\sigma_{j}^{(n)}+\beta u h_{i}^{(n)}\big),\label{d1}\\
	J_{ij}^{(n+1)}&=&J_{ij}^{(n)}\left(1-\frac{\tau}{\tau^{\prime}}\right)+\frac{\tau}{\tau^{\prime}}\sigma_{i}^{(n)}\sigma_{j}^{(n)}\tanh\beta.\label{eq:d2}
\end{eqnarray}


It has been shown numerically that if one selects external stimuli $\{\bb\xi^\mu\}_{\mu=1}^K$ by sampling uniformly the index $\mu$ from the index set $\{1,\dots, K\}$, the dynamics governed by~\eqref{d1}-\eqref{eq:d2}, yields a coupling matrix $\bb J$ that, for long times, tends to converge to Hebb's prescription~\citep{agliari2022pavlov}.\\

We now show that the analytical results obtained in Sec.~\ref{sec:continuous} hold also in the discrete case. Moreover, we compute the amplitude of the fluctuation around the steady state in the general case of non-uniformly presented patterns.\\
Let us focus on~\eqref{eq:d2}, we first analyze the situation where $\frac{\tau}{\tau '} < 1$ and the neurons are clamped to certain fixed values\footnote{Previous work only report analytical estimation in this case, see~\citep{agliari2022pavlov} Appendix A.2.} $\boldsymbol{\sigma}=\boldsymbol{\xi}^1$, then we generalize to the case where neuron values change.  
Similarly to what we have done in the continuous case let us define, for the sake of convenience, the following auxiliary parameters
\begin{equation*}
\epsilon := \frac{\tau}{\tau '}, \quad c:=\xi^1_i\xi^1_j\tanh\beta
\end{equation*}
 so that Eq.~\eqref{eq:d2}  can be written as 
\begin{equation}\label{fix}
    J_{ij}^{(n+1)}= 
  (1- \epsilon) J_{ij}^{(n)} + \epsilon c \, .
\end{equation}
Observing that no stochasticity is involved in the dynamics, for each $n$, we can replace the value of $J^{(n)}_{ij}$ with that from the previous iterations $n-1$. Proceed recursively, we obtain the following
\begin{align*}
    J_{ij}^{(n+1)} &=  (1- \epsilon) [  (1- \epsilon) J_{ij}^{(n-1)} + \epsilon c ] + \epsilon c \\&\ \ \vdots \\
    &= (1- \epsilon) ^n  J_{ij}^{(0)} + \epsilon c\sum_{k = 0 }^{n} (1- \epsilon)^k .
\end{align*}
Let us now observe that, since $J_{ij}^{(0)}$ is finite and $|1- \epsilon| < 1$ the first term on the right-hand side goes to zero as $n$ approaches infinity. The second term, instead, being a geometric series with a ratio $1- \epsilon < 1$, will converge to $\frac{1}{\epsilon}$.\\
Bringing everything together and substituting the explicit value of $c$, we get
\begin{equation}\label{ref:J_single_pattern}
    \lim_{n\to\infty }J_{ij}^{(n+1)}= \xi^1_i\xi^1_j\tanh\beta,
\end{equation}
which corresponds to the single pattern Hebbian prescription.
Let us remark that, the hyperbolic tangent in Eq.\eqref{ref:J_single_pattern} acts as a global attenuating factor that accounts for the thermal noise; for sufficiently low temperatures we have $\tanh\beta  \approx 1$.

If we have two patterns $K=2$, then the coupling matrix converges to the mixed Hebbian kernel 
\begin{equation}\label{eq:2patternshebbiankernel}
    J_{ij}:= \frac{1}{2} \left( \xi_i^1 \xi_j^1 + \xi_i^2 \xi_j^2 \right) \tanh \beta \, .\end{equation}
In Appendix~\ref{sec:appB}, we show how this result can be easily achieved by clamping neurons to pattern $ \bb  \xi^1$ during even iteration steps and to pattern $ \bb  \xi^2$ during odd steps. We also discussed how the same logic can be extended to the case with an arbitrary number of patterns cyclically presented to the network.

\subsection{Characterization of the steady state}\label{sec:prob}
Suppose stimuli are not presented in periodic cycles. In that case, we can only make statements about the probability distribution of the coupling matrix elements as the number of iteration steps goes to infinity.
Furthermore, in the real world, neural systems do not receive uniformly distributed external signals. In fact, certain concepts or stimuli will appear more frequently than others. Hence, it's relevant to ask if similar results are obtainable with a heterogeneous presentation of external stimuli. \\
To deal with this more realistic setting, now each pattern, denoted as $\boldsymbol{
\xi}^\mu$ with $\mu \in\{ 1, 2, \ldots, K\}$, has a certain probability $p_\mu$ of being presented at any iteration step.
In principle, this probability can vary with the iteration step index. Thus, the update rule for the coupling matrix is
\begin{equation}
J_{ij}^{(n+1)} = (1- \epsilon) J_{ij}^{(n)} + \epsilon c^{\mu(n)},
\end{equation}
where $c^{\mu(n)} := \xi_i^{\mu(n)} \xi_j^{\mu(n)} \tanh\beta$ corresponds to the pattern presented at step $n$, and $\mu(n)$ is now a random variable representing the index of the pattern chosen according to the probabilities $\{ p_\mu (n) \}_{\mu=1,\ldots,K}$.
Obviously, it always holds $\sum _\mu p_\mu (n) = 1$, regardless of $n$.\\
To analyze the expected behaviour of the coupling matrix over time, we compute the expected value w.r.t. the realization of the stochastic process $c^{\mu(n)}$ that we  denote with $\mathbb{E} [J_{ij}^{(n+1)}]$,
\begin{equation}
\mathbb{E}[J_{ij}^{(n+1)}] = (1- \epsilon) \mathbb{E}[J_{ij}^{(n)}] + \epsilon \mathbb{E}[c^{\mu(n)}].
\end{equation}
From this equation, we see that the expected contribution of a randomly chosen pattern at step $n$ is proportional to $\mathbb{E}[c^{\mu(n)}] = \sum_{\mu=1}^K p_{\mu}(n) c^{\mu (n)}$ with a coefficient equal to the time scales ratio.\\
Moreover, we are interested in the case where the pattern distribution does not vary with time so that $c^{\mu(n)} \equiv c_{\mu}$, $p_{\mu} (n) \equiv p_{\mu}$, for every $\mu$ and $n$, and $\mathbb{E}[c_{\mu}] = \sum_{\mu=1}^K p_{\mu} c^{\mu}$. Then, we can simplify the equation to:
\begin{equation}
\mathbb{E}[J_{ij}^{(n+1)}] = (1- \epsilon) \mathbb{E}[J_{ij}^{(n)}] + \epsilon \sum_{\mu=1}^K p_\mu c^{\mu}.
\end{equation}
Thus the expected value of the coupling matrix elements evolves towards the weighted average of the contributions from all $K$ patterns, with the weights given by their respective probabilities. This can be shown with recursive substitutions, as before. Instead, here we observe that, as $n$ approaches infinity, in the steady state $\mathbb{E}[J_{ij}^{(n+1)}] = \mathbb{E}[J_{ij}^{(n)}] = \mathbb{E}[J_{ij}^{(\infty)}]$, we easily obtain:
\begin{equation}
\mathbb{E}[J_{ij}^{(\infty)}] = \frac{\epsilon}{1 - (1- \epsilon)} \sum_{\mu=1}^K p_\mu c^\mu = \sum_{\mu=1}^K p_\mu c ^\mu.
\end{equation}
We underline that the expected value of $c^{\mu}$ is equal to the expected value of the coupling matrix element $J_{ij}^{(\infty)}$\footnote{To shorten the notation we have not displayed the neuron indexes in $c^{\mu} \equiv c^{\mu}_{ij}$. }, and in the low-temperature regime they coincide with the generalized Hebbian prescription
\begin{equation}\label{eq:gen_Hebb_matrix_element}
J_{ij}=\sum_{\mu =1 }^{K} p_\mu \xi^{\mu}_i \xi^{\mu}_j .
\end{equation}

To compute the variance, we start with the definition:
\begin{equation}
\text{Var}[J_{ij}^{(n+1)}] = \mathbb{E}[(J_{ij}^{(n+1)} - \mathbb{E}[J_{ij}^{(n+1)}])^2].
\end{equation}
Substituting the update rule for $J_{ij}^{(n+1)}$ into the variance formula and expanding, we get:
\begin{align}
\text{Var}[J_{ij}^{(n+1)}] &= \mathbb{E}\left[\left((1- \epsilon) J_{ij}^{(n)} + \epsilon c^{\mu(n)} - \mathbb{E}[(1- \epsilon) J_{ij}^{(n)} + \epsilon c^{\mu(n)}]\right)^2\right] \nonumber \\
&= \mathbb{E}\left[\left((1- \epsilon) (J_{ij}^{(n)} - \mathbb{E}[J_{ij}^{(n)}]) + \epsilon (c^{\mu(n)} - \mathbb{E}[c^{\mu(n)}])\right)^2\right]. \nonumber
\end{align}
We expand this expression and apply the linearity of expectation,
\begin{align*}
\text{Var}[J_{ij}^{(n+1)}] &= (1- \epsilon)^2 \text{Var}[J_{ij}^{(n)}] + \epsilon^2 \text{Var}[c^{\mu(n)}] \\
&\quad + 2(1-\epsilon)\epsilon \mathbb{E}[(J_{ij}^{(n)} - \mathbb{E}[J_{ij}^{(n)}])(c^{\mu(n)} - \mathbb{E}[c^{\mu(n)}])].
\end{align*}
Given that $J_{ij}^{(n)}$ and $c^{\mu(n)}$ are independent (assuming the pattern at step $n$ is independent of the coupling matrix's previous state), the last term vanishes.
\begin{equation}
\text{Var}[J_{ij}^{(n+1)}] = (1- \epsilon)^2 \text{Var}[J_{ij}^{(n)}] + \epsilon^2 \text{Var}[c^{\mu(n)}].
\end{equation}
The variance of $c^{\mu(n)}$ is equal to the expected value of the square of $c^{\mu(n)}$ minus the square of the expected value of $c^{\mu(n)}$.
\begin{equation}
\text{Var}[c^{\mu(n)}] = \sum_{\mu=1}^K p_\mu 
 \left( c^{\mu(n)} \right) ^2 - \left(\sum_{\mu=1}^K p_\mu c^{\mu(n)}\right)^2. \label{eq:vj}
\end{equation}
Under the assumption that the pattern distribution does not vary with time, as $n$ approaches infinity, since the system reaches a steady state, the variance of the coupling matrix elements stabilizes,\footnote{An alternative approach to reach the same result involves explicitly expressing  $J_{ij}^{(n+1)}$ as a function of the realizations $\{c^{\mu (i)} \} _{i=0,\dots, n}$ and computing its variance. Equation~\eqref{eq:vj} is derived by recognizing the geometric series and taking the limit $n \to \infty$.} and we obtain
\begin{equation} \label{eq:var_esplicit}
\text{Var}[J_{ij}^{(\infty)}] = \frac{\epsilon^2}{1 - (1- \epsilon)^2} \left[\sum_{\mu=1}^K p_\mu \left( c^\mu \right)^2 - \left(\sum_{\mu=1}^K p_\mu c^\mu\right)^2\right].
\end{equation}
Equivalently, we can leave implicit the variance of $c^{\mu}$ and simplify the initial factor.
\begin{equation} \label{eq:var_implicit}
\text{Var}[J_{ij}^{(\infty)}] = \frac{\epsilon}{2 - \epsilon} \text{Var}[c^{\mu}] = 
\left( \frac{2}{2 - \epsilon} - 1 \right) \text{Var}[c^{\mu}] 
\end{equation}
When the ratio between the neuron and the synaptic time scales goes to zero, the variance of the steady-state coupling matrix element goes to zero.
This is because $\epsilon$ regulates the system response for any single pattern presentation, the lower $\epsilon$ the lower the variation of $J_{ij}^{(\infty)}$ during an updating step. Thus, in the low $\epsilon$ regime, to make $J_{ij}^{(\infty)}$ significantly different from its mean, the network should be exposed to an atypical pattern sequence, e.g. the index random variable $\mu$ is uniform and we draw the same index multiple times. 
Actually, in this limit, we can Taylor expand Eq.~\eqref{eq:var_implicit} to get
\begin{equation}\label{eq:variance_taylor}
\text{Var}[J_{ij}^{(\infty)}] \approx \left( 1 + \frac{\epsilon}{2} + O(\epsilon^2) \right) \text{Var}[c^{\mu}] - \text{Var}[c^{\mu}] = \frac{\epsilon}{2} \text{Var}[c^{\mu}] + O(\epsilon^2) \text{Var}[c^{\mu}].
\end{equation}
Neglecting the terms of $O(\epsilon^2)$, we obtain 
\begin{equation} \label{eq:matrixelement_variance}
    \text{Var}[J_{ij}^{(\infty)}] 
    = \frac{\tanh^2\beta}{2} \frac{\tau}{\tau '} \left[ 1 - \left(  
\sum_{\mu =1 }^{K} p_\mu \xi^{\mu}_i \xi^{\mu}_j  \right)^2\right] \xrightarrow[\beta \to \infty]{}
    \frac{1}{2} \frac{\tau}{\tau '} \left[ 1 - \left(  
\sum_{\mu =1 }^{K} p_\mu \xi^{\mu}_i \xi^{\mu}_j  \right)^2\right], 
\end{equation}
where we used the fact that in the low-temperature limit $(c^{\mu})^2 = 1$. In Eq.~\ref{eq:matrixelement_variance}, we recognize the Hebbian matrix element~\eqref{eq:gen_Hebb_matrix_element}. 
In particular, the square root of~\eqref{eq:matrixelement_variance}, i.e.
 $\Delta_{ij}:=\sqrt{    \text{Var}[J_{ij}^{(\infty)}] }$ is the expectation of the displacement between $J_{ij}^{(\infty)}$ and the Hebbian prescription given in Eq.~\ref{eq:gen_Hebb_matrix_element}.

In conclusion, we obtain  from  Eq.~\ref{eq:matrixelement_variance}, the normalized Frobenius norm\footnote{Let us recall that, given a $n\times m$ matrix $\bb A$, the Frobenius norm is defined as $\lVert \bb A \rVert:=\sqrt{\sum_{i=1}^n\sum_{j=1}^m A_{ij}^2}$, the adjective \textit{normalized} comes from the fact that we have divided the value of the Frobenius norm by $\sqrt{m\times n}$.} 
\begin{equation}\label{eq:frob_distance}
     \lVert  \bb \Delta \lVert _F 
 = \frac{\tanh{\beta}}{N} \sqrt{ 
 \frac{1}{2} \frac{\tau}{\tau '} 
 \sum_{i,j=1}^N \left[ 1 - \left(  
\sum_{\mu =1 }^{K} p_\mu \xi^{\mu}_i \xi^{\mu}_j  \right)^2\right]} .
\end{equation}

\subsubsection*{Comparison with previous work}
The authors of~\citep{agliari2022pavlov} propose the following estimation for the fluctuation amplitude of the matrix elements 
\begin{equation}\label{eq:altro_paper_gen}
\tilde{\Delta}_{ij} \propto \sqrt{\frac 1 2\frac{\tau}{\tau'}\tanh^2\beta\left[1-\left(\sum_{\mu =1 }^{K} p_\mu \xi^{\mu}_i \xi^{\mu}_j \right)^2\right]+\left[\sum_{\mu =1 }^{K} p_\mu \xi^{\mu}_i \xi^{\mu}_j (1-\tanh\beta)\right]^2} \ .
\end{equation}
They only derive the low-temperature limit of this formula and to do so they make the ansatz for the evolution of the coupling matrix element $J_{ij} (t) =  p_\mu \xi^{\mu}_i \xi^{\mu}_j + \delta_{ij}(t)$, where the last term is a fluctuation contribution centred around zero that they assume can be written as $  \delta_{ij}(t) = 2 \tilde{\Delta} _{ij} \cos{ \left( t\sqrt{\frac{\tau}{\tau'}} + \phi \right) }$. They derive the dynamics of  $ \delta_{ij}(t)$ from the synaptic dynamics and finally find $\tilde{\Delta} _{ij}$ by imposing that the dynamics of $ \delta_{ij} ^2(t)$  is satisfied on average over a long time-scale.\footnote{We note that they write an expression for $2 \tilde{\Delta} _{ij}$, and the factor $2$ is arbitrary, hence we report here only the proportionality relation. \\ We have instead chosen the standard deviation as the natural scale for the fluctuation and found agreement in numerical experiments. Moreover the ansatz $\cos{t\sqrt{t \frac{\tau}{\tau'}}}$ is not supported by numerical evidences. Since the fluctuation is aperiodic, see e.g. Fig.~\ref{conv_families}.} In the low-temperature limit the amplitude of the fluctuation is proportional to $\sqrt{\epsilon} = \sqrt{\frac{\tau}{\tau '}}$, we also found the coefficient of this proportionality relation.\\
We observe that in the computation of variance, we don't find the term in \eqref{eq:altro_paper_gen} that does not depend on the time scale ratio.  \\

\section{Beyond Hebb's paradigm: the dreaming prescription}\label{sec:dreaming}
Until now, we have been working in the mean-field limit, the same approximation used on the standard Hopfield network in Amit, Gutfreund and  Sompolinsky (AGS) theory~\citep{amit1989}. 
Indeed, it is well-known that the Hebbian prescription (resulting in the celebrated Hopfield model~\citep{hopfield1982hopfield}) is not the optimal construction for the coupling matrix $\bb J$. In particular, from AGS theory~\citep{amit1989,CKS}, we see that the network can handle at most $K \sim 0.14 N$ patterns, while -- according to Gardner's theory -- spin-glass models with pairwise interactions can manage up to $K \sim 2N$ patterns (or $K \sim N$ for symmetric coupling matrices)~\citep{gardner1988optimal}. 
In particular, Hopfield model fails to retrieve stored information when the spurious states, corresponding to mixtures of memories, get exponentially more abundant than pure memories. 
Various approaches have been used to solve this problem, we cite \textit{unlearning algorithms} ~\citep{Kohonen-1984,Personnaz-JPhysLett1985, kanter1987associative, Vaas-PhysA1990,Plakhov-IEEE1992,FAB-NN2019}, which are based on the idea of increasing the energy of spurious state configurations to make them less stable. In particular, the unlearning algorithm of Plakhov and Semenov~\citep{Plakhov-IEEE1992} assumes that the synaptic matrix is updated by progressively removing the correlations between the total fields acting on the spins, rather than the correlations between spins directly.
 Inspired by this algorithm, we generalize the modeling discussed in previous sections to make it suitable for framing more performing, and biologically plausible, learning mechanisms.
 To this extent, we aim to adapt the dynamics~\eqref{eq:sigma_rad}-\eqref{eq:J_rad} so that the coupling matrix, by learning directly from the total fields acting on the spins, converges to the so-called Kanter and Sompolinsky projector ~\citep{kanter1987associative} which is given by
\begin{equation} \label{eq:dreamkernel}
J_{ij}^{(Dream)}:=\frac{1}{K}\sum_{\mu=1}^K\sum_{\nu=1}^{K}\xi_{i}^{\mu}{(C^{-1})_{\mu\nu}}\xi_{j}^{\nu}
\end{equation}
where
\begin{equation}
C_{\mu\nu}:=\frac{1}{N}\sum_{i=1}^{N}\xi_{i}^{\mu}\xi_{i}^{\nu},
\end{equation}
is the pattern correlation matrix. This network can handle a number of patterns $K\sim N$, thus saturating Gardner's theoretical bound.\\
The generalisation of Kanter and Sompolinsky's projector is the \textit{Reinforcement \& Removal} algorithm~\citep{FAB-NN2019}, also called the {\it dreaming prescription}; hence the label {\it Dream} in Eq.~\eqref{eq:dreamkernel}.
If we want the synaptic dynamics given in Eq.~\eqref{eq:J_rad} to converge to Eq.~\eqref{eq:dreamkernel}, we need to modify the way in which external stimuli act on the system. To illustrate this, let us notice that
\begin{equation}
J_{ij}^{(Dream)}=\frac{1}{K}\boldsymbol{\xi}_{i}^{T}{\boldsymbol{C}}^{-1}\boldsymbol{\xi}_{j}=\frac{1}{K}\left({\boldsymbol{C}^{-1/2}}\boldsymbol{\xi}_{i}\right)^{T}{\boldsymbol{C}^{-1/2}}\boldsymbol{\xi}_{j}=\frac{1}{K}\tilde{\boldsymbol{\xi}}_{i}^{T}\tilde{\boldsymbol{\xi}}_{j},\label{eq:dreaming}
\end{equation}
where $\boldsymbol{\xi}_{i}=(\xi_{i}^{1},\dots,\xi_{i}^{K})$, and $\tilde{\boldsymbol{\xi}}_{i}:= \boldsymbol{C}^{-1/2} \boldsymbol{\xi}_{i}$. This holds because the pattern correlation matrix is positive-definitive, thus the (principal) inverse square-root is well-defined. Now we see that, in order to recover the kernel~\eqref{eq:dreamkernel}, we need to apply stimuli which are aligned to the patterns $\tilde{\boldsymbol{\xi}}$. This modification of the stimuli cannot be directly learnt by the synaptic matrix $J_{ij}$ if we do not modify the synaptic dynamics too. 
In the synapses time-scale the process of forgetting spurious memories takes place.
Since we are not explicitly distinguishing between wakefulness and dreaming phases, our generalization can be seen as a daydreaming network~\citep{serricchio2024daydreaminghopfieldnetworkssurprising}, or even as a toy model for the hypnagogic state, since the network dynamics we are about to write is still affected by external stimuli~\citep{ghibellini2023hypnagogic}.\\
We let the synaptic matrix learn directly from the total field acting on the neurons, as in Plakhov's algorithm~\citep{Plakhov-IEEE1992}. This modification allows each coupling matrix element dynamics to depend on the local field acting on the neurons connected by it.

\begin{eqnarray}
\sigma_{i}^{(n+1)}&=&\tanh\left(\beta\sum_{i\neq j}^{N}J_{ij}^{(n)}\sigma_{j}^{(n)}+\beta
 u \tilde h_{i}^{(n)}\right):=\tanh(\beta H_{i}^{(n)}),\\
J_{ij}^{(n+1)}&=&J_{ij}^{(n)}\left(1-\frac{\tau}{\tau^{\prime}}\right)+\frac{\tau}{\tau^{\prime}}\frac{H_{j}^{(n)}}{V^{(n)}}\frac{H_{i}^{(n)}}{V^{(n)}}\tanh\beta. \label{eq:J_dream}
\end{eqnarray}
Here, $\tilde h_i$ is the external stimuli obtained by rotating the patterns according to the inverse square root of the correlation matrix, while $ H_i:=\sum_{i\neq j}^{N}J_{ij}^{(n)}\sigma_{j}^{(n)}+
 u \tilde h_{i}^{(n)}$ is the total field acting on the $i$-th spin. 
The normalisation factor $V^{(n)}$ is fixed as follows $V^{(n)}=\frac{\sqrt{N}}{\left\Vert \boldsymbol{H}^{(n)}\right\Vert}$
in such a way that the rescaled total field has norm $\left\Vert \frac{\boldsymbol{H}^{(n)}}{V^{(n)}}\right\Vert =\sqrt{N}$ that is the norm of a vector in an $N$-dimensional space whose every entry is either $+1$ or $-1$, and the neural configuration is one such vector. We also observe that normalisation allows the coupling matrix element to perceive bounded signals of different intensities, 
comparing Eq.~\eqref{eq:J_dream} with  Eq.~\eqref{eq:J_rad}, we see that in the latter the signals were made bounded by the hyperbolic tangent of Eq.~\eqref{eq:sigma_rad}. \\
Similar to the previous approach, one can once again prove the convergence of the dynamics to
\[ J_{ij}^{(\infty)} = \sum _ {\mu =1} ^ {K}  p_\mu \Tilde{\xi}^{\mu}_i \Tilde{\xi}^{\mu}_j \tanh{\beta} ,
\]
provided we are in our typical learning setting.\footnote{In the strong field regime $u\gg1$ and when the pattern field $\bb h$ changes sufficiently quickly (see the discussion around~\eqref{eq:effective_c_temporal_average}).} 
We remind that in this setting the neural time scale is well separated from the synaptic one,  and we assume that in the temporal window, regardless of $t$ and $\tau'$, neurons are aligned to each pattern $\bb \xi ^ \mu$ an amount of time that is proportional to the probabilities $\{ p_\mu \}_{\mu=1,\ldots, K}$.
In particular, if $\beta \gg 1$ and  $p_\mu = \frac{1}{K}$ for each pattern, we recover the Hebbian kernel~\eqref{eq:dreaming}.
%
We illustrate this point in our numerical experiments, presented in Sec.~\ref{sec:experiment}.

\section{Results and numerical evidences} \label{sec:experiment}
We have analytically characterized the model's behaviour in the case where the presented stimuli are non-uniformly distributed, with a certain pattern being more recurrent, and we also have modified the dynamical equations to converge to the dreaming recipe. In this section, we illustrate these results with numerical experiments. In particular, we will show the convergence of the synaptic matrix $\bb J^{(n)}$ to different Hebbian kernels $\bb J^{(final)}$, whose explicit expression depends on the scenario under consideration. Moreover, we will check the robustness of the convergence by studying the fluctuation amplitude around the expected convergence value~\eqref{eq:frob_distance}.

\begin{figure}	
			\includegraphics[clip,trim=0.02cm 0.09cm 0.075cm 0.122cm, width=\textwidth]{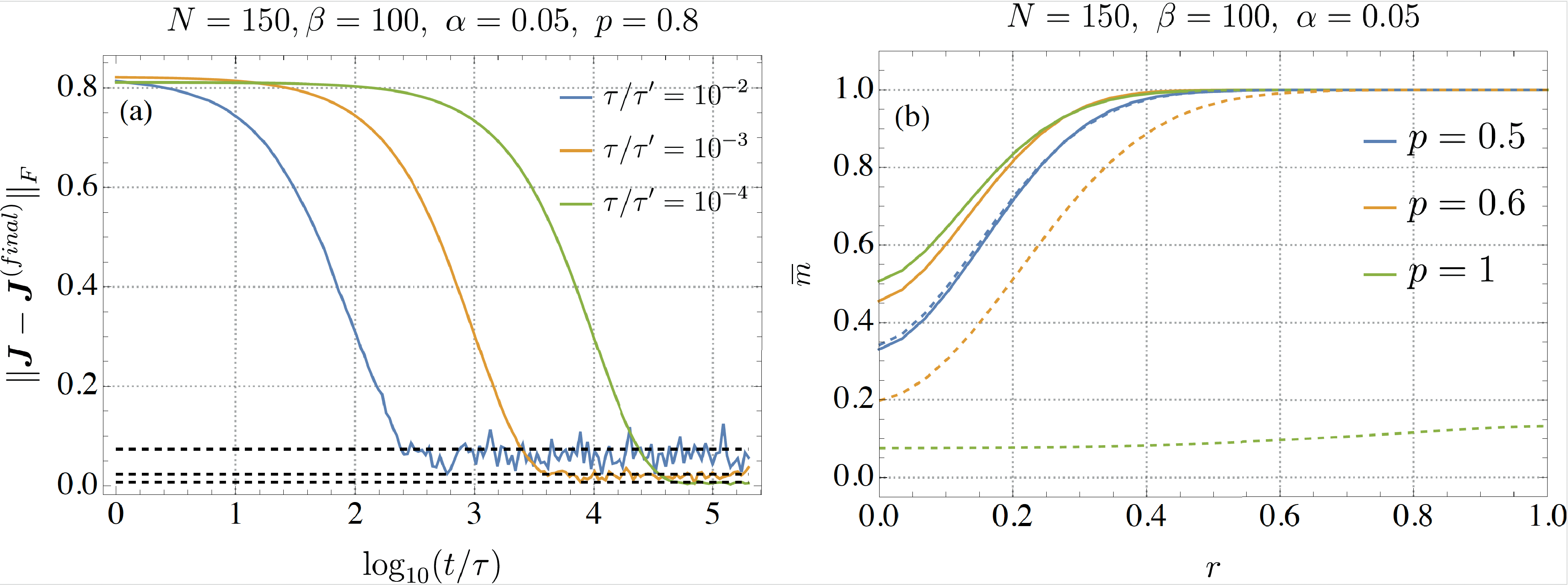}
\caption{{\bfseries Results related to the dynamics for binomially extracted patterns.} Left panel: Learning performances of the dynamics given by~\eqref{d1} and~\eqref{eq:d2}, quantified by the normalized Frobenius distance of the synaptic matrix w.r.t. the mixed Hebb's kernel~\eqref{eq:we_are_family} (i.e. $\bb J^{(final)}=\bb J^{(mixed)}$) for $\tau/\tau'=10^{-2},10^{-3},10^{-4}$. The asymptotes correspond to the values of the normalized Frobenius norm of $\boldsymbol \Delta$ given in Eq.~\eqref{eq:frob_distance} and adapted to the specific case in which the Hebbian kernel is $\boldsymbol{J}^{mixed}$.
Right panel: Retrieval capabilities as a function of the quality parameter $r$ for $p = 0.5, 0.6, 1$ and $M=100$.  The solid curves refer to the Mattis magnetizations of patterns belonging to the first
family, while dashed ones to the second family. The network parameters are $N = 150$, $u = 150$, $\beta =100$, $\alpha=0.05$,  for both the panels, with $\alpha$ being the storage capacity $\alpha=K/N$.}\label{conv_families}
\end{figure}

As a first experiment, we partition the set of $K$ patterns in two families $\{\bb \xi ^{1,\mu}\}_{\mu=1}^{K_1}$ and $\{\bb \xi ^{2,\mu}\}_{\mu=1}^{K_2}$, where $K=K_1+K_2$. We proceed by randomly selecting a pattern from the first family with a probability of $p$, and from the second family with a probability of $1-p$. Once the class is determined, the pattern is uniformly chosen from within that class and presented to the network. The pattern is extracted uniformly once the class is chosen. 
Since this procedure is semantic ({\it i.e.} it only involves index $\mu$) and the Hebbian schemes do not allow temporal correlation between the patterns ({\it i.e.} there is no correlation between patterns with different indices), it is quite natural to expect that the whole training procedure converges toward a matrix $J^{(final)}$ that takes the form of mixed Hebb interaction matrix

\begin{equation}\label{eq:we_are_family}
	J_{ij}^{(mixed)}:= p J_{ij}^{(1)}+(1-p)J_{ij}^{(2)},
\end{equation}
where
$$
J^{(a)}_{ij}:=\frac1{K_a} \sum_{\mu =1 }^{K_a}\xi^{a,\mu}_i \xi^{a,\mu}_j,
$$
for $a\in\{1,2\}$.\\
In order to study the convergence of the coupling matrix to the prescription~\eqref{eq:we_are_family}, we first analyze the evolution of the synaptic matrix $\bb J^{(n)}$ by computing, for each iteration, the normalized Frobenius norm of the difference between the synaptic matrix obtained from the numerical solution of equations~\eqref{d1}-\eqref{eq:d2} and the corresponding mixed Hebbian kernel $\mathbf J^{(mixed)}$. The convergence of the coupling matrix is reported in the left panel of Fig.\ref{conv_families} where learning performances of the dynamics~\eqref{d1}-\eqref{eq:d2} is shown for different choices of the ratio $\tau/\tau'$. More precisely, the plot is obtained by selecting the pattern family according to a Bernoulli experiment with a fixed parameter $p$, subsequently presenting to the network a uniformly selected pattern from the chosen family. As expected, the different choice of the ratio $\tau/\tau'$ affects both the convergence rate of the algorithm and the amplitude of the fluctuations around the mean values of the normalized Frobenius distance Eq.~\ref{eq:frob_distance} as $t\to\infty$. 
It is then clear that the closer the timescales are, the less accurate the retrieval will be. Only when the timescales are completely separated, the normalized Frobenius distance converges to zero (see Eq.~\eqref{eq:frob_distance}).
The convergence of the coupling matrix to the prescription~\eqref{eq:we_are_family} is reported in Fig.~\ref{conv_families} (left panel) for various values of the ratio $\tau/\tau'$, and for families with the same number of patterns, i.e. $K_1=K_2=K/2$. 
 In Fig.~\ref{conv_families} (right panel) we report the retrieval capabilities of the resulting model.
 To test the retrieval capabilities of the network, for each pattern, we sample $M$ noisy versions of it by randomly flipping a percentage of bits equal to $(1-r)/2$, with $r\in(0,1]$; in this context, $r$ can be interpreted as the quality of the noisy pattern generated. These examples are used as initial condition for the dynamics. We let the spins evolve according to Eq.~\eqref{d1} for $5$ iterations, with no stimuli (i.e $h_i=0$) and with the synaptic matrix fixed and equal to $\boldsymbol{J}=\boldsymbol{J}^{mixed}$.
The final neural configuration $\boldsymbol{\sigma}^{(\infty)}$ depends on the initial configuration which, in turn, depends on both the original pattern $\boldsymbol{\xi}^{a,\mu}$ and on the realization of the noise injection process that we have just described and that creates $M$ noisy patterns for each $\mu\in\{1,\dots,K_a\}$ that we can label with the new subscript $\bar a$.
Thus, $\boldsymbol{\sigma}^{(\infty)} \equiv \boldsymbol{\sigma}^{(\infty)}_{\bar a}(\boldsymbol{\xi}^{a,\mu})$ with $\bar a \in \{1,\dots,M\},a\in \{1,2\},\mu\in\{1,\dots,K_a\}$. 
As a performance metric we use Mattis magnetization~\citep{camilli2022inference} which is defined as follows
 \begin{equation}
 m(\boldsymbol{\sigma}^1,\boldsymbol{\sigma}^2) :=\frac{\sum_{i=1}^N\sigma^{1}_i\sigma^{2}_i}{N}
 \end{equation}
 where $\boldsymbol{\sigma}^1,\boldsymbol{\sigma}^2\in \Sigma_N$. We observe that $m\in[-1,1]$ and if the two configurations are equal then their Mattis magnetization is equal to $1$  on the contrary if they are orthogonal then the Mattis magnetization is zero. We evaluate the Mattis magnetization between each of the $M$ final neural configurations and the corresponding pattern and then we take the average over the index $\bar a$ to get $m_{\mu,a}=\frac 1 M \sum_{\bar a =1}^M m(\boldsymbol{\sigma}^{(\infty)}_{\bar a}(\boldsymbol\xi^{a,\mu}),\boldsymbol\xi^{a,\mu}))$. We repeat this algorithm for every pattern, and we take the average also over $\mu$ index inside each family to get $\bar{m}_a=\frac{1}{K_i}\sum_{\mu=1}^{K_i}m_{\mu,a}$, for $a=1,2$. The Mattis magnetizations for the two families are reported as a function of $r$ and for various values of the probability $p$. For $p=1/2$, we recover a learning kernel which is proportional to the Hebbian kernel of the Hopfield model, and the magnetizations follow the same behaviour. For $p\neq1$, we have a separation between the two families instead. In particular, the higher $p$, the higher the average Mattis magnetization of the first family of patterns, meaning that the corresponding patterns are associated with more stable configurations for the energy function. These results can be easily extended to the case of 
 linear combinations of $\Letteranuova$ Hebbian kernels
 with individual probabilities $p_\letteranuova$:
\begin{equation}\label{generalJ}
J_{ij}^{(final)}=\sum_{\letteranuova=1}^\Letteranuova p_\letteranuova J^{(\letteranuova)}_{ij},
\end{equation} 
with $\Letteranuova$ being the number of families, and clearly $p_1+\dots+p_\Letteranuova=1$.

%
\par
\begin{figure}	
		\centering
		\includegraphics[width=\textwidth]{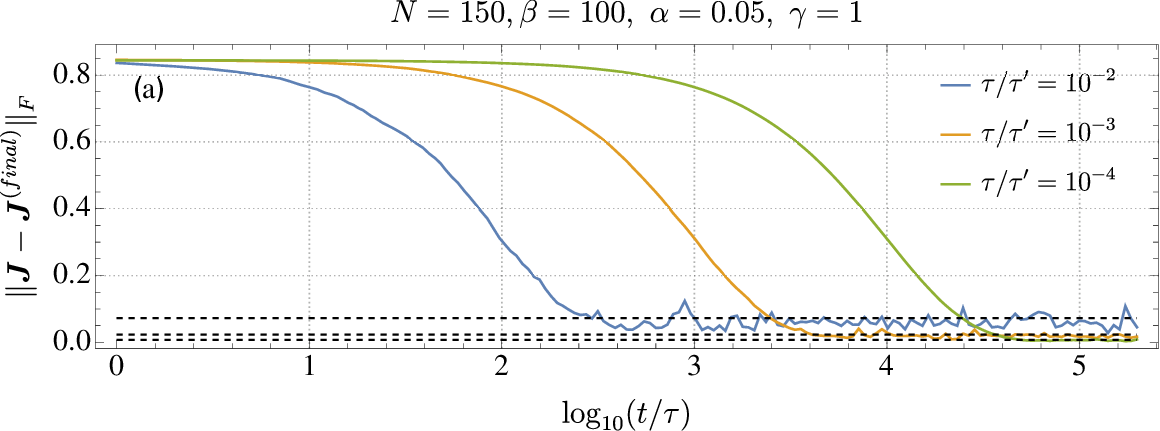}
  \\
        \includegraphics[clip,trim=0.2cm 0.15cm 0.015cm 0.07cm,width=0.98\textwidth]{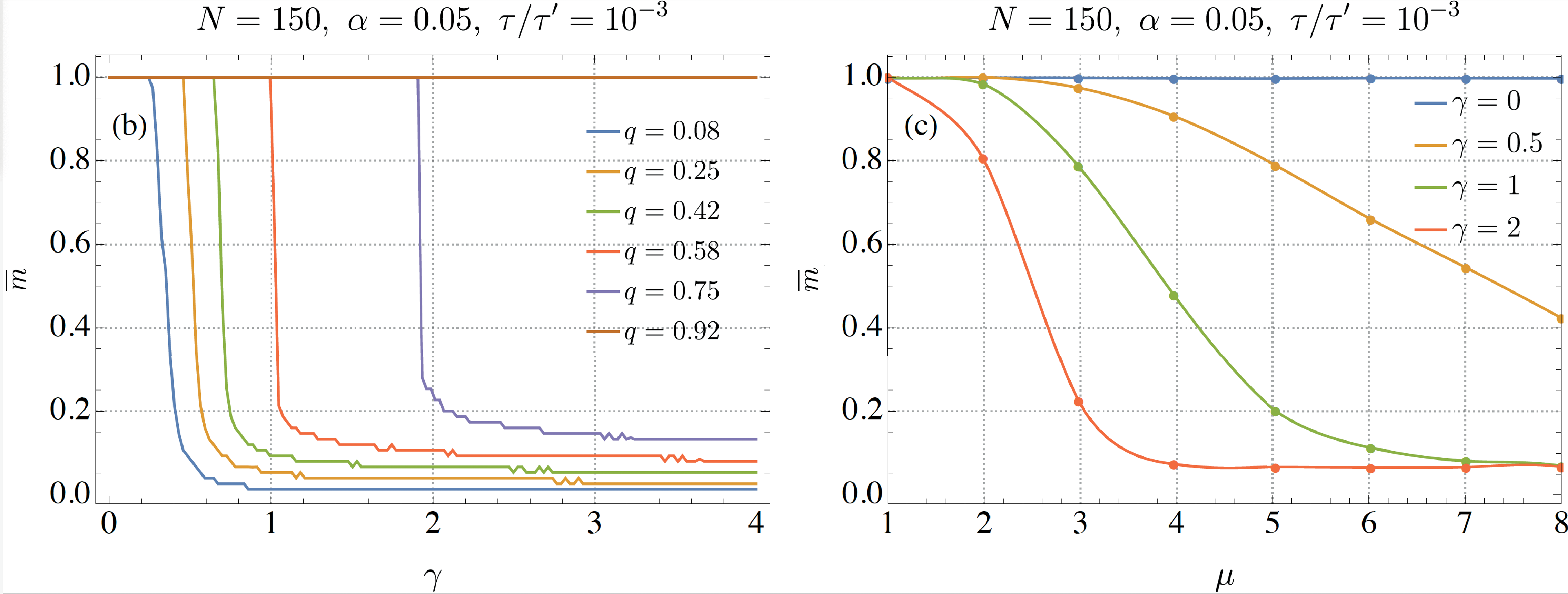}
        \caption{Top panel: Convergence of the synaptic matrix towards the power-law Hebbian kernel (i.e. $\bb J^{(final)}=\bb J^{(power)}$) for the ratio $\tau/\tau'=10^{-2},10^{-3},10^{-4}$. The curves are obtained by averaging the results of $1500$ different realizations of the retrieval process, and the horizontal dashed lines correspond to the expected normalized Frobenius distance, Eq.~\ref{eq:frob_distance}. Bottom left panel: Mattis magnetization curves as a function of the exponent $\gamma$ for $q=0.08,0.25,\dots,0.92$. Bottom right panel: Mattis magnetization as a function of the pattern index for $\gamma=0,0.5,1,2$. The network parameters are $N=150$, $\alpha=0.05$, $\beta =100$ for both panels. The network parameters are $N = 150$, $u = 150$, $\beta =100$, $\alpha=0.05$,  for all the panels,  with $\alpha$ being the storage capacity $\alpha=K/N$.}
        \label{hebb_pow1}
\end{figure}

\bigskip
For our second experiment, we explore the setting, where, instead of families of patterns chosen according to a discrete probability distribution, we have $K$ patterns that we present to the network with different probabilities
$p_\mu$, with $\mu = 1, \, ..., \, K$. 
We chose the pattern, that acts as the external stimulus,  based on a scale-free distribution. This choice is particularly interesting since many real-world distributions adhere to power laws. In other words, the index $\mu$ is chosen based on the following probability assignment:
\begin{equation}\label{eq:power_law_prob}
p_{\mu}= \frac{c_{K,\gamma}}{\mu^\gamma}, \quad \mu=1,\dots,K,
\end{equation}
where $c_{K,\gamma}$ is a normalization constant depending on the number of stored patterns and on the coefficient $\gamma$. This scenario corresponds to the special case where we have a single pattern per family and $K$ total families. So by using the formula~\eqref{generalJ} with $K_a=1,\Omega =K,p_a=\frac{c_{K,\gamma}}{a^{\gamma}}$ we get
\begin{equation}\label{eq:power_law_hebb}
J^{(power)}_{ij} = c_{K,\gamma}\sum_{\mu=1}^K \xi^\mu_i \frac{1}{\mu^\gamma} \xi^\mu_j.
\end{equation}
As in the previous case, we study the convergence of the synaptic matrix towards the power-law Hebbian kernel by computing at each iteration the normalized Frobenius norm of the difference between $\bb J$ and $\bb J^{(power)}$ (see Fig.~\ref{hebb_pow1}, top panel.

Then, we analyze the retrieval capabilities of the resulting model as a function of both the scaling exponent $\gamma$ (Fig.~\ref{hebb_pow1} left panel) and the index $\mu$  related to patterns (Fig.~\ref{hebb_pow1} right panel).\\
We expect that, for a given $\gamma$, there will be some patterns that the network retrieves correctly and others that are not retrieved. This is because the retrieval depends on the probability given by Eq.~\eqref{eq:power_law_prob}: given a pattern $\boldsymbol \xi ^\mu$, the higher the label $\mu$ the more rarely it will be presented to the network and the more difficult it will be for the network to retrieve it. 
To test our claims we select patterns at random, we sample the initial neural configurations with quality $r$ w.r.t. the patterns (just as we did before), we run the neural dynamics given in Eq.~\eqref{d1} with no stimuli and with the synaptic matrix fixed and equal to Eq.~\eqref{eq:power_law_hebb} and we plot the distribution of the resulting magnetizations $\bar m= m(\boldsymbol{\sigma}^{(\infty)}(\boldsymbol{\xi}^\mu),\boldsymbol{\xi}^\mu)$.
We see a peak at $\bar m= 1$, that corresponds to the patterns that are correctly retrieved, then we locate the remaining probability mass at $\bar m  \gtrsim 0$ because, for the rest of the patterns, the retrieval fails.
The results of this experiment for various $\gamma$ are shown in Fig.~\ref{hebb_pow1}, bottom left panel. There, we plot the quantiles $q$ of the distribution of magnetization values for increasing values of  $\gamma$, each of them corresponds to a different simulation. The so obtained curves start from $\bar m= 1$, at $\gamma = 0$, because in this case, the uniform distribution is recovered and the parameters of the network correspond to the retrieval phase of the Hopfield model with load $\alpha \ll \alpha _c$, where $\alpha _c\approx 0.14 $ is the critical load of the Hopfield model.
Then, $\bar m$ will eventually drop because some patterns are not retrieved. This drop occurs for lower $\gamma s$ for the lower quantiles because the lower $\gamma$ is, the more probability mass will be close to $\bar m  = 1$ \footnote{For all of these experiments, we also varied the network parameters ({\it i.e.} the network size $N$, the storage capacity $\alpha$, the thermal noise $\beta$) both for the binomially and power-law extracted patterns, and we found that the training performances are analogous to those of the pure Hebbian case.}.

\begin{figure}[!ht]
	\includegraphics[width=\textwidth]{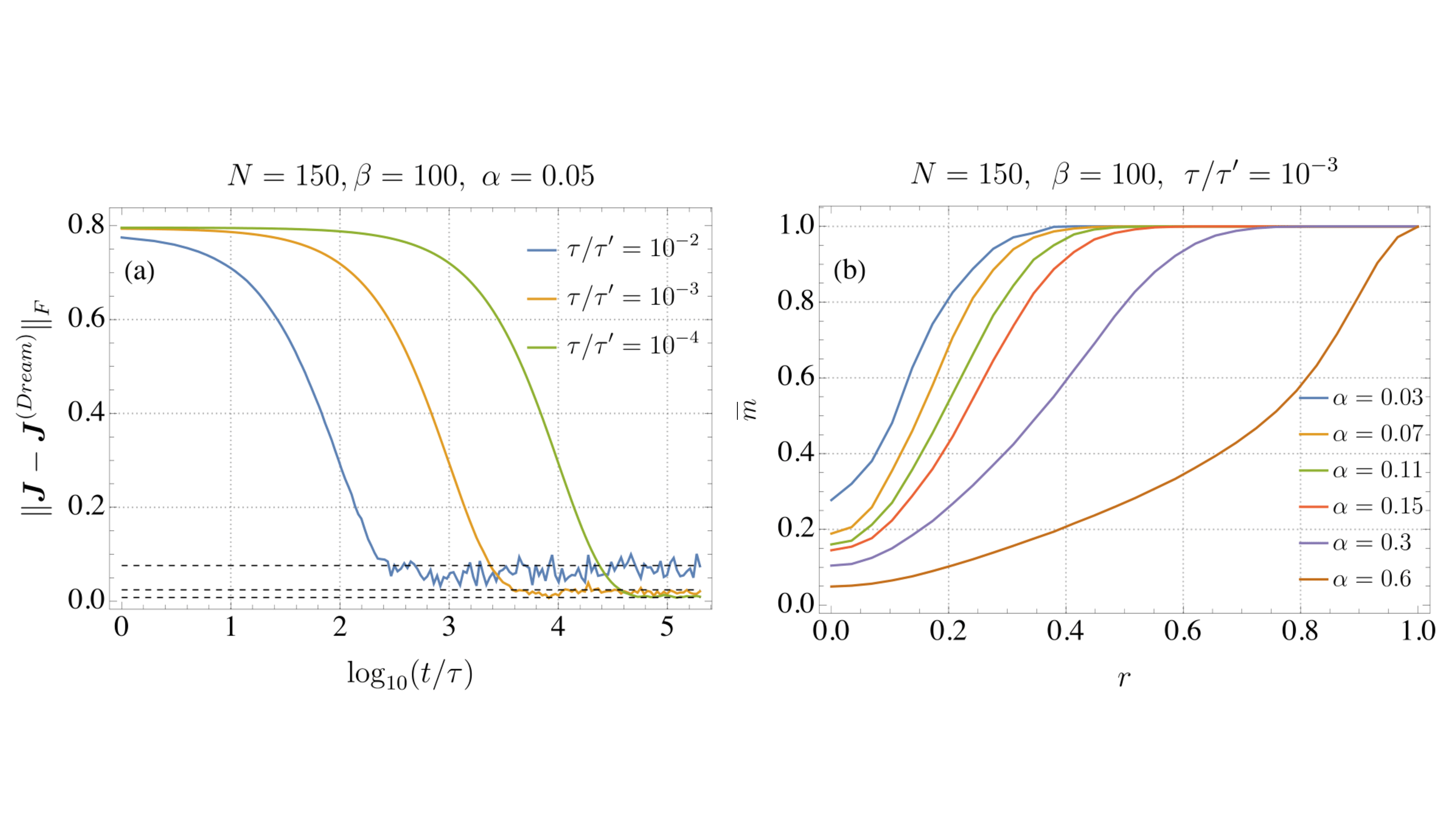}
\caption{{\bfseries Results for the modified synaptic dynamics that leads to the dreaming prescription.} Left panel: normalized Frobenius distance, Eq.~\ref{eq:frob_distance}, of the synaptic matrix w.r.t. the dreaming kernel for $\tau/\tau'=10^{-2},10^{-3},10^{-4}$. Right Panel: Mean Mattis magnetizations as a function of the quality parameter $r$ for $\alpha=0.03,\, 0.07,\ldots, \,0.6$. 
The other network parameters are $N = 150$, $u = 150$, and $\beta =100$  for both the panels. }\label{fig:dreaming_res} 
\end{figure}

Moving to a numerical analysis of the dreaming prescription, in Fig.~\ref{fig:dreaming_res}, top panel, we plot the distance between the coupling matrix and the dreaming prescription~\eqref{eq:dreaming} as a function of the evolution time, for different choices of the ratio $\tau/\tau'$, finding that the algorithm converges, as expected. We observe that the norm of the difference between the dreaming and the numerical coupling matrices is given by~\eqref{eq:frob_distance}.
We also observe that the dynamics follows~\eqref{eq:eqdiff_k1_sol}, as it happened in all the precedent numerical scenarios. 
In Fig.~\ref{fig:dreaming_res}, bottom left panel, we present the results concerning the system's retrieval capabilities after training. This scenario is achieved by keeping the resultant coupling matrix fixed, and the results are plotted as a function of the quality $r$ for different values of the storage capacity $\alpha=K/N$. Evidently, the model can manage a larger information content compared to the Hebbian case reported in~\citep{agliari2022pavlov}. Additionally, at a sufficiently low storage capacity, the system handles low-quality examples more effectively than the standard Hopfield model. It is noteworthy that when comparing Fig. 3 in~\citep{agliari2022pavlov} (representing the Hebbian case) with Fig.~\ref{fig:dreaming_res} (bottom-left panel, representing the dreaming kernel), the latter model exhibits greater stability against perturbations in the initial conditions for retrieval dynamics, especially at higher values of the quality parameter $r$. \\
This implies that the model, when equipped with the dreaming kernel, features broader attraction basins related to the stored patterns. As a result, the retrieval quality increases, especially at very high storage capacities where the Hebbian counterpart fails to achieve effective retrieval. This result is in perfect agreement with the analysis for the attraction basins performed in~\citep{FAB-NN2019}.\\
Finally, we emphasize that when the load of the network is below the higher critical storage capacity, the results are analogous to the pure Hebbian case. Whilst, above this threshold, the network properly converges to the dreaming kernel.

\section{Conclusions}\label{sec:end}
In conclusion, this paper explores a mathematical model that is able to capture the fundamental features of both Pavlovian classical conditioning and Hebbian learning. \\
We have discussed how a system, where both the couplings and the spins are considered thermodynamics variables can memorize a set of patterns that are presented through a magnetic field regulated by an amplification factor. The so-obtained mean-field dynamics is then solved by integrating the equations under appropriate conditions. We have discussed how the model is able to reproduce Pavlov classical conditioning mechanisms, and we have analytically proven that the Hebbian coupling matrix is obtained through this dynamics, the variance of the learning stochastic process has been also computed.
We illustrated the analytical results in the setting in which the stimuli are divided into two families and the network sees one family of stimuli more frequently than the other. In such scenarios, the coupling matrix converges to a weighted element-wise sum of the coupling matrices of the two families, with the weights determined by the probabilities of selecting each family.\\
Finally, in the spirit of describing biological networks within our framework, we have shown that the model can describe neural networks that sleep. Pavlovian learning rules are indeed compatible with the dreaming mechanism used in the associative neural network literature. 
In this case, the dynamics of the synapses allow our model's coupling matrix to converge towards the so-called dreaming kernel. The network not only handles more patterns compared to its purely Hebbian counterpart but also exhibits enhanced retrieval quality and resilience against initial condition perturbations, especially at higher storage capacities. 
We have discussed this setting, emphasizing that all analytical considerations regarding convergence remain valid in this case. \\


\section*{Acknowledgments}
The authors acknowledge Adriano Barra, Elena Agliari, and Alberto Fachechi for the precious discussions about this work.\newline
D.L. acknowledges INdAM-GNFM and C.N.R. (National Research Council). \\

\appendix
\section{The master equation derivation of the coupled dynamics} \label{sec:appA}

A more formal approach to derive the evolution equations~\eqref{d1}-\eqref{eq:d2} for the expectation values of $\sigma_{i}$ and $J_{ij}$ over time makes use of the master equation related to their stochastic process~\citep{agliari2022pavlov}. We sketch this formulation in this appendix and remand to the paper for the full formal discussion. \\
We start by introducing $w_{N,\beta}(\bb \sigma, \bb J\to \bb \sigma^{\prime}, \bb J^{\prime})$, the transition rate between two system states $(\bb\sigma,\bb J)$ and $(\bb\sigma^{\prime},\bb J^{\prime})$, and let $\mc P_{N, \beta}(\bb \sigma,\bb J,t)$ denote the probability of the system being in the configuration $(\bb\sigma,\bb J)$ at time $t$. \\
The master equation governing the time evolution of the probability $\mc P_{N, \beta}(\bb \sigma,\bb J,t)$ is given by:

\begin{align}
\frac{d}{dt}\mc P_{N, \beta}(\bb \sigma,\bb J,t) = & \sum_{\bb\sigma^{\prime}\in \Sigma_{N}}\sum_{\bb J^{\prime}\in \mathcal{J}_{N}}\Bigg[\mc P_{N, \beta}(\bb \sigma^{\prime},\bb J^{\prime},t)w_{N,\beta}(\bb J^{\prime},\bb \sigma^{\prime}\to \bb J,\bb \sigma) \notag \\
& - P_{N, \beta}(\bb \sigma,\bb J,t)w_{N,\beta}(\bb J,\bb \sigma\to \bb J^{\prime},\bb \sigma^{\prime})\Bigg]
\label{master}
\end{align}
Reference~\citep{agliari2022pavlov} proceeds assuming that neural activities and synaptic weights are updated simultaneously to factorize the transition rate $w( \bb\sigma^{\prime}, \bb J^{\prime}\to \bb \sigma,\bb J)$ as
\begin{equation}
w_{N,\beta}( \bb\sigma^{\prime}, \bb J^{\prime}\to \bb \sigma,\bb J)=\prod_{l,m}\frac{1}{2}\left[1+\tanh\left(\beta J_{lm}\sigma_{l}^{\prime}\sigma_{m}^{\prime}\right)\right]\prod_{i}\frac{1}{2}\left[1+\sigma_{i}\tanh\left(\beta\sum_{j}J_{ij}^{\prime}\sigma_{j}^{\prime}+\beta h_{i}\right) \right].
\end{equation}
The transition rate is thus expressed as a product of single-variable conditional probabilities. \\
\medskip \par
Focusing on neural dynamics first, the mean neural activity is defined as  
\begin{equation} \label{eq:barsigma}
\langle\sigma_{i}\rangle=\sum_{\bb \sigma\in \Sigma_{N}}\sum_{\bb J\in \mathcal{J}_{N}}\mc P_{N,\beta}(\bb \sigma,\bb J,t) \, \sigma_{i}, ~~ \textrm{for} ~~ i=1,...,N
\end{equation}
and set the typical timescale for its dynamics to be $\tau$. \\ 
Multiplying the master equation~\eqref{master} by $\sigma_{i}$, and summing over all possible neural and synaptic configurations, we arrive at the following stochastic process:
\begin{eqnarray} \label{eq:process1}
\tau\frac{ d\langle\sigma_{i}\rangle}{dt} = &+& \sum_{\bb J,\bb J^{\prime}\in\mathcal{J}_{N}}\sum_{\bb \sigma,\bb \sigma^{\prime}\in \Sigma_{N}}\sigma_{i}[\mc P_{N,\beta}(\bb \sigma^{\prime},\bb J^{\prime},t)w_{N,\beta}(\bb J^{\prime},\bb\sigma^{\prime}\to \bb J,\bb \sigma) ]\\ \nonumber
&-& \sum_{\bb J,\bb J^{\prime}\in\mathcal{J}_{N}}\sum_{\bb \sigma,\bb \sigma^{\prime}\in \Sigma_{N}}\sigma_{i}[\mc P_{N,\beta}(\bb \sigma,\bb J,t)w_{N,\beta}(\bb J,\bb \sigma\to \bb J^{\prime},\bb\sigma^{\prime})].
\end{eqnarray}
It is observed that, by definition,
\begin{equation}
\sum_{\bb \sigma^{\prime}\in \Sigma_{N}}\sum_{\bb J^{\prime}\in \mathcal{J}_{N}}w_{N,\beta}(\bb J,\bb \sigma\to \bb J^{\prime},\bb \sigma^{\prime})=1 .
\end{equation}
This identity can be used to simplify~\eqref{eq:process1},
\begin{equation}\label{exps}
\tau \frac{d \langle\sigma_{i}\rangle } {dt}=-\langle\sigma_{i}\rangle +\sum_{\bb J,\bb J^{\prime}\in\mathcal{J}_{N}}\sum_{\bb \sigma,\bb \sigma^{\prime}\in \Sigma_{N}}\sigma_{i}\mc P_{N,\beta}(\bb \sigma^{\prime},\bb J^{\prime},t)w_{N,\beta}(\bb J^{\prime},\bb\sigma^{\prime}\to \bb J,\bb \sigma)
\end{equation}
which, in turn, is equivalent to
\begin{equation}\label{preMFT}
\tau\frac{d  \langle\sigma_{i}\rangle}{dt}=- \langle\sigma_{i}\rangle +\left\langle \tanh\left(\beta\sum_{j}J_{ij}\sigma_{j}+\beta h_{i}\right)\right\rangle .
\end{equation}
By applying the mean field approximation, equation (\ref{preMFT}) can be recast as
\begin{equation} \label{eq:sigma_radnew}
\tau\frac{d \langle\sigma_{i}\rangle }{dt} = -\langle\sigma_{i}\rangle+\tanh\Big(\beta\sum_{j} \Big \langle J_{ij} \Big \rangle \langle\sigma_{j}\rangle+\beta h_{i}\Big).
\end{equation}
It is noted that, while at the individual neuron level, neural firing is typically described by Langevin equations~\citep{Tuckwell1988}, at this coarse-grained treatment, the evolution of the mean firing rate is governed by an ODE: the random terms are averaged out, but the noise $\beta$ is still a central quantity. Indeed, for $\beta \to 0$ neither the post-synaptic potentials nor the external stimuli can be perceived, and it is the term $-\langle \sigma_{i} \rangle$ in~\eqref{eq:sigma_radnew}  that ensures the mean firing rate to return to zero.\\
\medskip \par
Moving to the synaptic dynamics, its typical timescale can be assumed to be \(\tau' > \tau\), and the dynamical evolution of the average of synaptic weights, defined by
\[
\langle J_{ij}\rangle=\sum_{\bb\sigma\in \Sigma_{N}}\sum_{\bb J\in\mathcal{J}_{N}}\mc P_{N,\beta}(\bb\sigma,\bb J,t)J_{ij}
\]
is obtained by multiplying the master equation by \(J_{_{ij}}\) and summing over all possible neural configurations and synaptic matrices. \begin{align}
\tau' \frac{d \langle J_{ij}\rangle }{dt} = &+ \sum_{\bb J,\bb J^{\prime}\in\mathcal{J}_{N}}\sum_{\bb \sigma,\bb \sigma^{\prime}\in \Sigma_{N}}J_{ij}\left[\mc P_{N,\beta}(\bb \sigma^{\prime},\bb J^{\prime},t)w_{N,\beta}(\bb J^{\prime},\bb\sigma^{\prime}\to \bb J,\bb \sigma)\right] \nonumber \\
&- \sum_{\bb J,\bb J^{\prime}\in\mathcal{J}_{N}}\sum_{\bb \sigma,\bb \sigma^{\prime}\in \Sigma_{N}}J_{ij}\left[\mc P_{N,\beta}(\bb \sigma,\bb J,t)w_{N,\beta}(\bb J,\bb \sigma\to \bb J^{\prime},\bb\sigma^{\prime})\right],
\end{align}
which simplifies to
\[
\tau' \frac{d \langle J_{ij}\rangle }{dt}=- \langle J_{ij}\rangle +\sum_{\bb J^{\prime}\in\mathcal{J}_{N}}\sum_{\bb \sigma^{\prime}\in \Sigma_{N}}\mc P_{N,\beta}(\bb \sigma^{\prime},\bb J^{\prime},t)\tanh\left(\beta\right)\sigma_{i}^{\prime}\sigma_{j}^{\prime},
\]
which under the mean field approximation, leads to
\begin{equation} 
\tau' \frac{d\langle J_{ij}\rangle }{dt}=-\langle J_{ij}\rangle +\tanh\left(\beta\right) \langle \sigma_{i} \rangle \langle \sigma_{j} \rangle.\label{eq:J_radnew} 
\end{equation}
Equations~\eqref{eq:sigma_rad} and~\eqref{eq:J_radnew} constitute the system of coupled ODEs, and are equivalent to~\eqref{d1} and~\eqref{eq:d2}. 

\section{The statistical mechanics derivation of the system equilibria} \label{sec:app0}
In this appendix, we derive equations~\eqref{eq:a} and ~\eqref{eq:b}. \\
Starting from equation~\eqref{eq:a}, 
\begin{equation} 
\begin{split}
\frac{\partial\log {{Z}_{N}(\beta, \bb t, \bb T)}}{\partial T_{ij}}\Big\vert_{\bb T=0, \bb t =0}&=\frac{1}{Z_N(\beta,t,\bb t)}\frac{\partial Z_{N}(\beta,t,\bb t)}{\partial T_{ij}}\Big\vert_{\bb T=0, \bb t =0}\\&=\frac{\sum_{\boldsymbol \sigma}\exp(\beta u\sum_{k}h_k\sigma_k )\sinh(\beta \sigma_i\sigma_j)\prod_{\substack{k<l\\(k,l)\ne (i,j)}}\cosh(\beta \sigma_k\sigma_l)}{\sum_{\boldsymbol \sigma}\exp(\beta u\sum_{k}h_k\sigma_k )\prod_{k<l}\cosh(\beta \sigma_k\sigma_l)}\\&=\frac{\sum_{\boldsymbol \sigma}\exp(\beta u\sum_{k}h_k\sigma_k )\tanh(\beta \sigma_i\sigma_j)\prod_{k<l}\cosh(\beta \sigma_k\sigma_l)}{\sum_{\boldsymbol \sigma}\exp(\beta u\sum_{k}h_k\sigma_k )\prod_{k<l}\cosh(\beta \sigma_k\sigma_l)}\\&=\langle\tanh(\beta \sigma_i\sigma_j)\rangle=\langle \sigma_i\sigma_j\rangle\tanh(\beta) \ .
\end{split}
\end{equation}
Here, in the last step, we used the fact that \(\tanh (\sigma x ) = \sigma \tanh x\) for \(\sigma=\pm1\).

Moving to equation \eqref{eq:b}, we first explicit the partition function for the neural subsystem, which sees couplings as parameters,
\begin{equation}\label{eq:starting_J}
{Z}_{N}(\beta, \bb t, \bb T; \bb J)=\sum_{\bb \sigma}\exp\Big(\beta\sum_{i<j}^N J_{ij}\sigma_{i}\sigma_{j}+\beta u \sum_{i=1}^N h_{i}\sigma_{i}+\sum_{i=1}^N t_{i}\sigma_{i}+\sum_{i<j}^NT_{ij}J_{ij}\Big).
\end{equation}
We now compute the neural expectation,
\begin{align}
&\frac{\partial\log {{Z}_{N}(\beta, \bb t, \bb T ; \bb J)}}{\partial t_i}\Big\vert_{\bb T=0, \bb t =0}=\frac{1}{Z_{N}(\beta,0,0)}\sum_{\boldsymbol{\sigma}}\prod_{k=1}^N\exp\left[\beta\sigma_k \Big( \sum_{l < k} J_{kl} \sigma_l 
+uh_k\Big) \right]\sigma_i\nonumber \\
& \approx \frac{\prod_{k\ne i}2\cosh\left[\beta\Big( \sum_lJ_{kl}\langle \sigma_l \rangle +uh_k\Big)\right]2\sinh\left[\beta\Big(
\sum_lJ_{il}\langle \sigma_l \rangle +uh_i\Big)\right]}{\prod_{k=1}^N2\cosh\left[\beta\Big( 
\sum_lJ_{kl}\langle \sigma_l \rangle +uh_k\Big)\right]} \label{eq:mf} \\
&=\tanh \Big( \beta \sum_{l\ne i}J_{il} \langle \sigma_l \rangle +\beta u h_i \Big) \nonumber 
\end{align}
We have used the shorthand $  \langle \sigma_l \rangle  =  \langle \sigma_l \rangle _{ \bb J \text{fixed}}$ to underline we are considering the couplings as parameters. We obtain Eq.~\eqref{eq:b} and the approximation accounts for the mean-field assumption, which in this case reads as 
\begin{align*}
\sigma_i \sigma_j &= (\sigma_i - \langle \sigma_i \rangle)(\sigma_j - \langle \sigma_j \rangle) + \sigma_i \langle \sigma_j \rangle + \langle \sigma_i \rangle \sigma_j - \langle \sigma_i \rangle \langle \sigma_j \rangle \\
&\approx \sigma_i \langle \sigma_j \rangle + \langle \sigma_i \rangle \sigma_j - \langle \sigma_i \rangle \langle \sigma_j \rangle\, . \quad \text{(neglecting fluctuations)}
\end{align*}
In the computation~\eqref{eq:mf}, the last term, $\langle \sigma_i \rangle \langle \sigma_j \rangle$, is not explicitly written because it cancels out with the corresponding term in the partition function.
.\\

\section{Analytical derivation of the Hebbian kernel from Pavlovian dynamics}\label{sec:appB}
In Section~\ref{sec:discrete} of the main text we proved the convergence of the discrete dynamics to the Hebbian kernel in the simple case where the neurons are clamped to a fixed pattern. Here we will analyze the more realistic case in which we let the neurons change. In this appendix, we focus on the unrealistic case where the patterns are presented cyclically to the network. \\
In the case of two patterns the neurons are clamped to $\boldsymbol\xi^1$ when the iteration step is even and to $\boldsymbol\xi^2$ when the iteration step is odd.

To simplify the notation, let us denote by $c_1:=\xi_i^1 \xi_j^1 \tanh\beta$, $c_2:=\xi_i^2 \xi_j^2 \tanh\beta$ and $\epsilon:= \frac{\tau}{\tau '}$. The dynamics of the coupling matrix can then be written as
\begin{equation}
    \begin{cases}
        J_{ij}^{(n+1)}= 
        (1- \epsilon) J_{ij}^{(n)} + \epsilon    c_1  \quad \text{if } n \equiv 0 \pmod{2}\\
        J_{ij}^{(n+1)}= (1- \epsilon) J_{ij}^{(n)} + \epsilon    c_2  \quad \text{if } n \equiv 1 \pmod{2} \, .
    \end{cases}
\end{equation}
Following the same approach used in Section~\ref{sec:discrete}, let us express $J_{ij}^{(n+1)}$ in terms of its previous values while considering the alternating influence of $c_1$ and $c_2$.
To simplify, we consider the sequence of updates as pairs, each pair covering a full cycle of the two patterns. Starting from an odd iteration step, 
\begin{align*}
J_{ij}^{(n+2)} &= (1- \epsilon) [ (1- \epsilon) J_{ij}^{(n)} + \epsilon   c_2 ] + \epsilon   c_1 \\
&= (1- \epsilon)^2 J_{ij}^{(n)} +  \epsilon [(1- \epsilon) c_2 + c_1].
\end{align*}
After $2n$ steps (or $n$ cycles), the coupling matrix evolves as:
\begin{equation}
J_{ij}^{(2n)} = (1- \epsilon)^{2n} J_{ij}^{(0)} +  \epsilon \left[ c_1 + (1- \epsilon) c_2 \right] \sum_{k = 0 }^{n-1} (1- \epsilon)^{2k}.
\end{equation}
The sum in the last term is a geometric series with ratio $(1- \epsilon)^2$, which converges, as $n$ approaches infinity to $\frac{1}{2\epsilon - \epsilon^2}$. Therefore we get 
\begin{equation}
\lim_{n\to\infty}J_{ij}^{(2n)} = \frac{1}{2 - \epsilon} \left[   c_1 +  (1- \epsilon) c_2 \right]
\label{eq:cycle_update}
\end{equation}
that, when the timescales are well-separated (i.e. $\epsilon \rightarrow 0$), is exactly mixed Hebbian kernel defined in ~\eqref{eq:2patternshebbiankernel}

Similarly, we compute the limiting coupling matrix, but starting, from an even iteration step. By denoting with $J_{ij,\texttt{even}}^{(\infty)}$ the ``even update'' convergence matrix and with $J_{ij,\texttt{odd}}^{(\infty)}$ the ``odd update'' convergence matrix, we obtain
\begin{align*}
\lim_{n\to\infty}J_{ij}^{(n)} &= \frac{1}{2} \left( J_{ij,\texttt{even}}^{(\infty)} + J_{ij,\texttt{odd}}^{(\infty)} \right) \\
&= \frac{1}{2}  \frac{1}{2 - \epsilon} \left[ c_1 + (1- \epsilon) c_2 + c_2 + (1- \epsilon) c_1 \right] \\
&= \frac{1}{2}  \frac{1}{2 - \epsilon}  \left[ c_1 \left( 2- \epsilon \right) + c_2  \left( 2- \epsilon \right) \right] \\
&= \frac{1}{2} \left( c_1 + c_2 \right) .
\end{align*}

We are now ready to extend this conclusion to an arbitrary number of patterns. Let's consider a scenario where we have $K$ patterns, denoted by $\{ \boldsymbol\xi^\mu \}_{\mu=1, \dots K}$, and the system alternates among these patterns in a sequential manner over time.

For each pattern $\mu$, let $c_\mu := \xi_i^\mu \xi_j^\mu \tanh\beta$. The dynamics of the coupling matrix when the system is exposed to these patterns in sequence can be generalized as follows:

\begin{equation}
J_{ij}^{(n+1)}=
\begin{cases}
(1- \epsilon) J_{ij}^{(n)} + \epsilon c_1 & \quad \text{if } n \equiv 0 \pmod{K}\\
(1- \epsilon) J_{ij}^{(n)} + \epsilon c_2 & \quad \text{if } n \equiv 1 \pmod{K} \\
\quad \vdots \\
(1- \epsilon) J_{ij}^{(n)} + \epsilon c_M & \quad \text{if } n \equiv K-1 \pmod{K}.
\end{cases}
\end{equation}

Following a similar approach as with two patterns, we consider the updates over $K$ cycles to capture the influence of all patterns. After $Kn$ steps (or $n$ complete cycles through all $K$ patterns), the coupling matrix evolves as:

\begin{equation}
J_{ij}^{(Kn)} = (1- \epsilon)^{Kn} J_{ij}^{(0)} + \epsilon \left[ \sum_{\mu=1}^{K} c_\mu \left( 1- \epsilon \right) ^{\mu -1} \right] \sum_{\eta = 0 }^{n-1} (1- \epsilon)^{K \eta}.
\end{equation}
The sum in the last term, $\sum_{\eta = 0 }^{n-1} (1- \epsilon)^{K \eta }$, is a geometric series with ratio $(1- \epsilon)^K$ which converges, as $n$ approaches infinity, to  $\frac{1}{1 - (1- \epsilon)^K}$. 
Therefore we can write
\begin{equation} \label{eq:generalonecycles}
\lim_{n\to\infty}J_{ij}^{(Kn)} = \frac{\epsilon}{1 - (1- \epsilon)^K} \sum_{\mu=1}^{K} c_\mu \left( 1- \epsilon \right) ^{\mu -1}= \frac{1}{K} \sum_{\mu=1}^{K} \xi_i^\mu \xi_j^\mu \tanh\beta,
\end{equation}
where in the last equality we used the fact that the first-order Taylor expansion for $\frac{\epsilon}{1 - (1- \epsilon)^K}$ simplifies to the normalization factor $\frac{1}{K}$, and that neglecting terms with the factor $\epsilon$ in the round brackets of the sum, we can further simplify this expression in the limit as $\epsilon \rightarrow 0$, which corresponds to well-separated timescales.

More rigorously, we can sum over all the possible $K$ cycle realizations, each of which starts from a different pattern $\boldsymbol\xi ^ \mu $ and finishes with the same pattern.
We can compute
\begin{equation}
\lim_{n\to\infty}J_{ij}^{(Kn)} = \frac{1}{K}  \frac{\epsilon}{1 - (1- \epsilon)^K} \sum_{\mu=1}^{K} c_\mu  \left[ 
 \sum_{\eta=1}^{K} \left(1 - \epsilon \right) ^ {\eta-1} \right] = \frac{1}{K}  \frac{\epsilon}{1 - (1- \epsilon)^K} \sum_{\mu=1}^{K} c_\mu  \frac{1 - (1-\epsilon)^K}{1-(1-\epsilon)}
\end{equation}
to reveal that Eq.\eqref{eq:generalonecycles} is an exact result.\\
Thus, we have demonstrated that, in the infinite time limit and for a system that sequentially alternates among $M$ patterns, the coupling matrix converges to the Hebbian prescription.

\newpage


\end{document}